\documentclass[journal=jacsat,manuscript=article]{achemso}
\setkeys{acs}{articletitle = true,doi = true}
\usepackage{amssymb,amsmath,bm}
\usepackage[usenames,dvipsnames]{xcolor}
\usepackage[bookmarks=false,colorlinks]{hyperref}
  \hypersetup{linkcolor=black,citecolor=black,filecolor=black,urlcolor=black}
\usepackage{multirow}
\usepackage{graphicx}
\usepackage{booktabs}
\usepackage{chemformula} 
\usepackage{tabularx}
\usepackage{multirow}
\usepackage{enumitem}
\usepackage{array}
\setlist{leftmargin=*,noitemsep}

\newcommand{\rev}[1]{\textcolor{black}{{#1}}}

\newcommand{\supf}{\textcolor{Black}{Figure S}}
\newcommand{\supt}{\textcolor{Black}{Table S}}

\newcommand*{\citen}{}% generate error, if `\citen` is already in use. source: https://tex.stackexchange.com/questions/94178/temporarily-disable-superscript-in-citation
\DeclareRobustCommand*{\citen}[1]{%
  \begingroup
    \romannumeral-`\x % remove space at the beginning of \setcitestyle
    \setcitestyle{numbers}%
    \cite{#1}%
  \endgroup
}



\author{Alexandru B.\ Georgescu}
\altaffiliation{Contributed equally to this work.}
\affiliation[Northwestern University]
{Department of Materials Science and Engineering, Northwestern University, Evanston, IL, USA}

\author{Peiwen Ren}
\altaffiliation{Contributed equally to this work.}
\affiliation[Northwestern University]
{Department of Materials Science and Engineering, Northwestern University, Evanston, IL, USA}

\author{Aubrey R.\ Toland}
\affiliation[MIT]{Department of Materials Science and Engineering, Massachusetts Institute of Technology, Cambridge, MA, USA}

\author{Shengtong Zhang}
\affiliation[Northwestern Industrial]{Department of Industrial Engineering and Management Sciences, Northwestern University, Evanston, IL, USA}

\author{Kyle D.\ Miller}
\affiliation[Northwestern University]
{Department of Materials Science and Engineering, Northwestern University, Evanston, IL, USA}

\author{Daniel Apley}
\affiliation[Northwestern Industrial]{Department of Industrial Engineering and Management Sciences, Northwestern University, Evanston, IL, USA}

\author{Elsa A.\ Olivetti}
\affiliation[MIT]{Department of Materials Science and Engineering, Massachusetts Institute of Technology, Cambridge, MA, USA}

\author{Nicholas Wagner}
\affiliation[Northwestern University]
{Department of Materials Science and Engineering, Northwestern University, Evanston, IL, USA}

\author{James M. Rondinelli}
\email{jrondinelli@northwestern.edu}
\affiliation[Northwestern University]
{Department of Materials Science and Engineering, Northwestern University, Evanston, IL, USA}

\title{Database, Features, and Machine Learning Model to Identify Thermally Driven Metal-Insulator Transition Compounds}

\begin{document}

\clearpage

\begin{abstract}
Metal-insulator transition (MIT) compounds are materials that may exhibit insulating or metallic behavior, depending on the physical conditions, and are of immense fundamental interest owing to their potential applications in emerging microelectronics. An important subset of MIT materials are those with a transition driven by temperature. The number of thermally driven MIT materials, however, is scarce, which makes delineating these compounds from those that are exclusively insulating or metallic challenging. Most research that addresses thermal MITs is limited by the domain knowledge of the scientists to a subset of MIT materials, and is often focused on a limited subset of possible features. Here, using a combination of domain knowledge and natural language processing (NLP) searches, we have built a material database comprising thermally driven MITs, as well as metals and insulators with similar chemical composition and stoichiometries to the MIT compounds. We featurized this dataset using a wide variety of compositional, structural, and energetic descriptors, including two MIT relevant energy scales, the estimated Hubbard interaction and the charge transfer energy, as well as the structure-bond-stress metric referred to as the global-instability index (GII). We then performed supervised classification on this dataset, constructing three electronic-state classifiers: metal \emph{vs} non-metal (M), insulator \emph{vs} non-insulator (I), and MIT \emph{vs} non-MIT (T). This classification allows us to identify new features separating MIT materials from non-MIT materials. These include the 2D feature space consisting of the average deviation of the covalent radius and the range of the Mendeleev number. We discuss the relationship of these atomic features to the physical interactions underlying MITs in the rare-earth nickelate family. We then elaborate on other features (GII and Ewald energy), and examine how they affect the classification of binary vanadium and titanium oxides. Last, we implement an online version of the  classifiers, enabling quick probabilistic class predictions  by uploading a crystallographic structure file. The broad accessibility of our database, newly identified features, and user-friendly classifier models, will aid in accelerating the discovery of MIT materials.
\end{abstract}
\maketitle
\sloppy
\newpage

\section{Introduction}
Metal-insulator transition (MIT) materials undergo an electronic phase change from a metallic to an insulating state as a function of applied external conditions, i.e., 
temperature, pressure, or doping. The transition is typically discerned through optical and/or transport 
measurements.\cite{Imada/Fujimori/Tokura:1998} 
Predicting whether a material is prone to undergo an MIT or not is an ongoing research 
area\cite{PhysRevLett.121.245701, Vargas-Hernandez2018, Dong2019} of high technological importance. MIT materials  
may deliver new ``steep slope'' transistors that operate at very low voltage for beyond-Boltzmann-based 
computation\cite{Shukla2015,Brahlek2017} or function as ``smart'' components in 
thermochromic windows.\cite{Cui2018}. To that end, there is strong interest in discovering new materials exhibiting MITs with improved properties.

\begin{figure*}[t]
\centering
\includegraphics[width=0.75\textwidth]{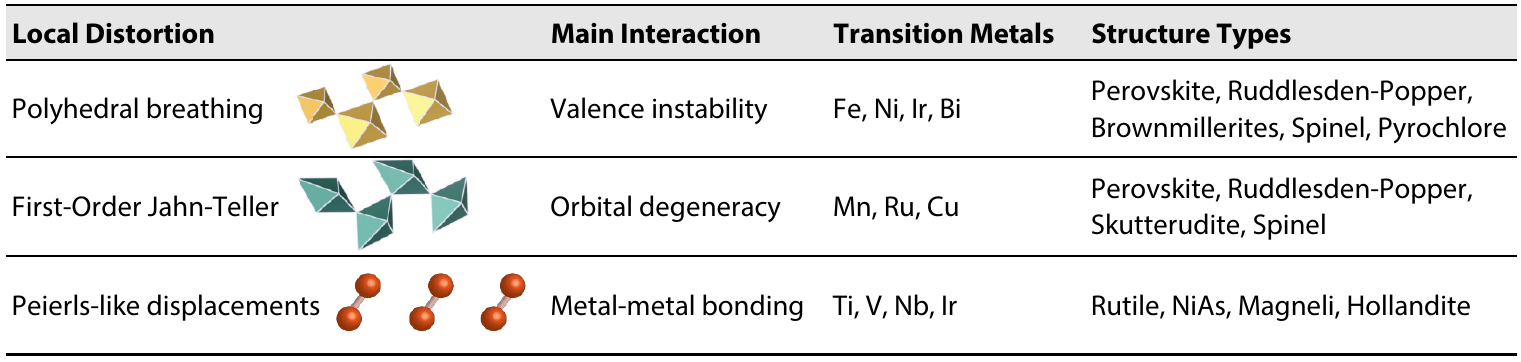}
\caption{Relationship between atomic distortions and the physical interactions driving MITs in transition metal compounds comprised of diverse chemistries and structure types.}
\label{fig:mechanisms}
\end{figure*}

Rational design of MIT materials, however, has proven to be difficult.
One reason for this is that the electronic phase transition may arise from a variety of (possibly competing) 
mechanisms (see \autoref{fig:mechanisms} for a non-exhaustive list).
The transition mechanism is often the subject of intense debate. The discussions are often limited to 
select chemistries and crystal structures\cite{Hiroi2015}. The transition is often characterized by certain microscopic physical observables that differentiate the metallic from the insulating state, which are then treated as variable order parameters.
In cases where the order parameters are known, 
there is often ambiguity over whether the transition is driven by the electronic or  
lattice order parameters, which hinders subsequent control and optimization of the transition characteristics, as well as over the role of the other different  structural and electronic modes in tuning the relative energies that characterize the transition\cite{wang2020featureless,Georgescu14434,Dominguez2020}.
\rev{While significant progress has been made recently in disentangling the electronic and lattice degrees of freedom in the low-temperature insulating state\cite{landscapes}, understanding the temperature dependence of the electronic properties of a material, and particularly whether a material with an insulating state at $T=0$\,K will transition to a metallic state, or vice versa, as the temperature is raised in a reliable way, is beyond the currently available high-throughput methods available in the field.}

Progress in the synthesis of high-quality materials, novel characterization methodologies, and advances in the quantum-mechanical modeling and theory of electron correlations has led to the recognition that subtle details in the crystal and local structure are indeed essential to describing MITs.\cite{PhysRevMaterials.4.104401,PhysRevMaterials.3.095003,GEORGESCU2021107991,landscapes} 
Displacive distortions to the size, shape, and connectivity of the basic metal-oxygen polyhedra or shortening of metal-metal distances in transition metal compounds are common in a number of these materials, which adopt different crystal structures (\autoref{fig:mechanisms}). 
This is best exemplified in the thermal MIT in VO$_{2}$, which 
appears to be described by a Mott-assisted Peierls transition, rather than exclusively 
Mott-Hubbard or Peierls-type physics\cite{Hiroi2015, Jager2017, Lee2018}. 
More recently, the perovskite oxide rare-earth nickelate family\cite{landscapes,Dominguez2020,Nick,Georgescu14434,Ghosez,Oleg,PhysRevLett.112.106404, Berciu,PhysRevMaterials.1.024410,Liao9515,PhysRevMaterials.1.024410} 
and the Ruddlesden-Popper ruthenate Ca$_2$RuO$_4$ \cite{landscapes,CRO, CRO2, CRO3, CRO4, CRO5} have been the subject of intense study, with both theoretical and experimental work focused on identifying and understanding the MIT mechanism to enable phase control. Improvements in sample quality have allowed for new metal-insulator transitions to be discovered even in previously known materials \cite{CMO}.  Although most MIT materials are oxides, there is an increasing amount of work focused at the discovery of materials with anions different from oxygen\cite{mixed,mixed2}.

\begin{figure}
    \centering
    \includegraphics[width=0.65\textwidth]{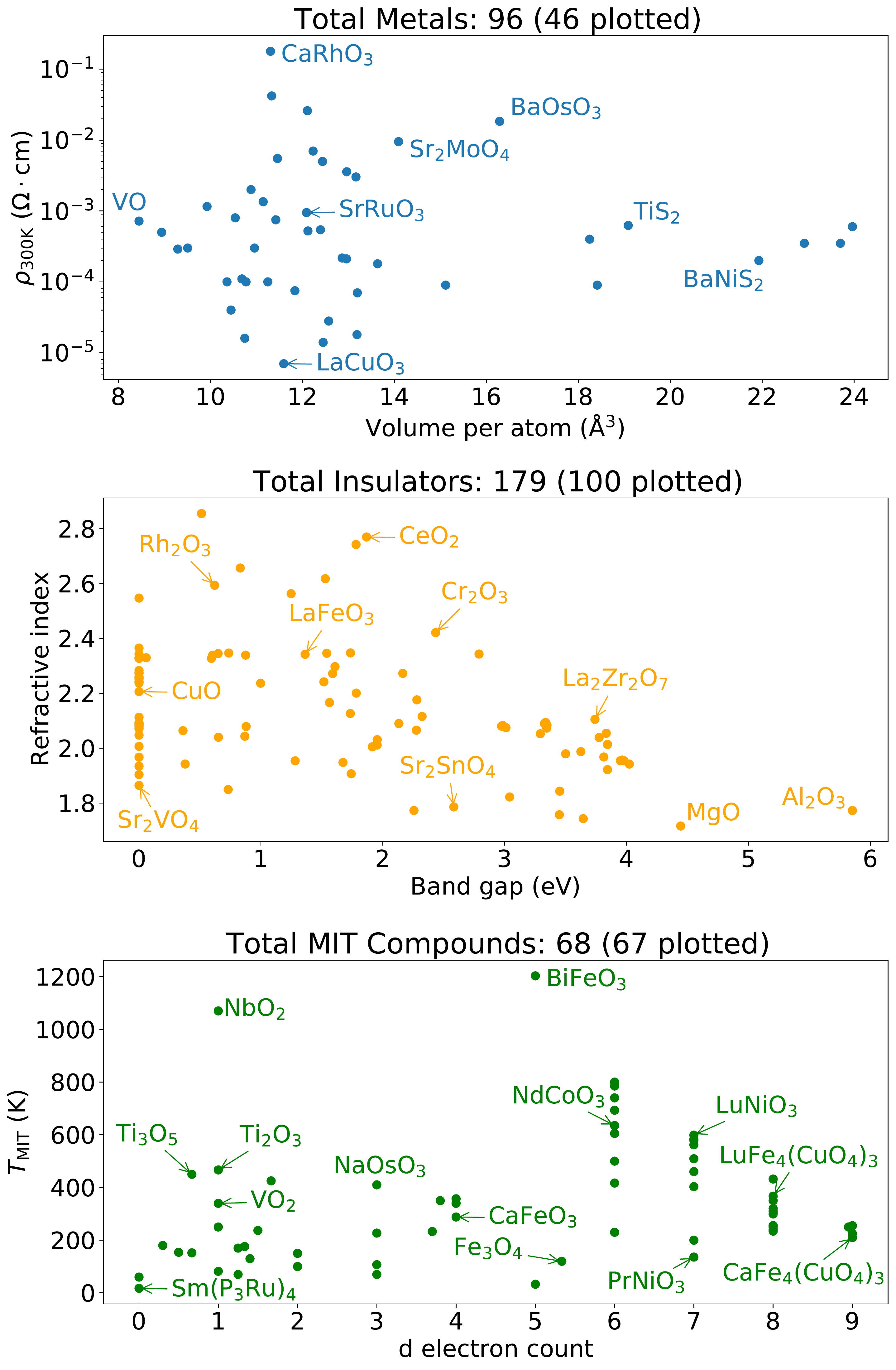}
    \caption{%
    Visualization of a sample of the database, comprising  (upper) metals, (center) insulators, and 
    (lower) thermal metal-insulator transition materials, in a 2D space of features commonly used in each electronic class.
    Not all compounds are visualized since sometimes the required data cannot be obtained (e.g., missing band gap values from the Materials Project database).
    The metals exhibit a broad range of resistivity values at 300\,K, and no clustering of the materials occurs although the vast majority exhibit $\rho<10^{-2}$\,$\Omega\cdot\mathrm{cm}$.
    The DFT calculated band gap energies, obtained from the Materials Project, and calculated refractive indices, following Shannon method,\cite{shannon_empirical_2016} are inversely proportional to each other.\cite{PhysRevMaterials.3.044602}
    (Band gap values of 0\,eV for some insulators are due to DFT underestimation.)
    Most MIT compounds exhibit transition temperatures $T_\mathrm{MIT}$ below 600\,K independent of d electron count. 
    (Note that the non-integer d electron count for some MIT compounds is a result of the averaging of oxidation states over several atomic sites.)
    }
    \label{fig:summary}
\end{figure}

Despite this progress, there are still less than a hundred inorganic materials which exhibit thermally-driven MITs (\autoref{fig:summary}). Further, despite their scarcity, before this work there has been no standard library of MIT materials available to the general scientific audience, which has slowed down the study of known MIT materials, and most likely reduced the rate at which new materials are discovered.
Experimental databases of synthesized and predicted inorganic 
materials, e.g., the ICSD\cite{Hellenbrandt2004} or SpringerMaterials\cite{SpringerMaterials} lack the necessary information to assign an electrical conductivity class labels: always  metallic, always insulating, or exhibiting a thermal MIT.
Although high-throughput first-principles databases exist, e.g., Materials Project\cite{Jain2013}, OQMD\cite{Saal2013MaterialsOQMD}, and AFLOW\cite{Curtarolo2012}, the methods used to compute the data often omit essential  microscopic interactions and corrections to standard DFT exchange-correlation functionals that could capture the MIT physics. The theory used (density functional theory) is also  $T=0$\,K theory. 
While these databases have been used to successfully build machine-learning classifiers for separating metals and insulators, as differentiated at a band-theory level (see \supt2 of the Supporting Information, SI), they often do not include the relevant physics to describe MIT materials families.  
Particularly, they do not model in sufficient detail the effect of electron-electron interactions that are crucial to understanding the opening of the band gap in correlated materials. 
High-throughput DFT without the use of appropriate microscopic models\cite{Varignon2019} or corrections specific to correlated materials can lead to the incorrect classification of MIT materials and some insulators as metals. 
For example, the insulator LaTiO$_3$ and the MIT material NdNiO$_3$ are both listed as having  0\,eV band gaps in Materials Project. Indeed, \autoref{fig:summary} shows that a large number of materials---including insulators---would be classified as metals based on the simulated 0\,eV band gap from Materials Project \rev{providing a poor starting point for any further classification}.
The combination of materials scarcity and inaccurate descriptions of the ground state are among the main difficulties in building machine-learning models for  discovering and understanding correlated materials.
Furthermore, limited efforts have focused on identifying whether any of these materials possess \emph{thermally-driven} MITs at finite temperatures, as the first-principles calculations often only report 0\,K data. 
\rev{Current theoretical methods are often insufficient to understand the complex temperature-dependent electron-lattice interplay leading to an MIT\cite{landscapes,Varignon2019}.}

Here, we resolve these limitations and build a database of experimentally confirmed temperature-driven MIT compounds through a combination of domain-knowledge and natural language processing (NLP).
We then augment this database with  structurally and compositionally related materials that are exclusively metallic or insulating. 
We featurize the complete dataset using atomic, electronic, and structural descriptors along with MIT-specific features, including an unscreened Coulomb interaction through an estimated Hubbard $U$ energy, $U_\mathrm{est}$, an estimated charge transfer energy, $\Delta_0$, and the global-instability index (GII).
After training multiple supervised learning models for the three classification  models, i.e., the metal \emph{vs} non-metal (M), insulator \emph{vs} non-insulator (I), and MIT \emph{vs} non-MIT (T) classification tasks, we identify new features whose interplay separate MIT materials from non-MIT materials.
Analysis of the SHAP scores of the T-model led us to identify two previously unappreciated descriptors that offer significant class separation without requiring sophisticated computational techniques:
($i$) average deviation of the covalent radius (ADCR), a feature that describes the relative size difference among the elements comprising a compound, and 
($ii$)  a compositional feature, called the range of the Mendeleev number. 
The GII and Ewald energy are also identified as important features.
We then examine the role these features play in the MITs exhibited by binary vanadates, titanates, and complex rare-earth nickelates.
Finally, we describe an online tool comprised of the three binary classifiers, which 
enables a user to upload a crystal structure file to obtain three probabilities of it being identified as a metal, insulator, or MIT compound.

\section{Methods}
Feature-based supervised learning involves data acquisition, feature engineering, and model building. 
The main result of our data acquisition is a database containing 343 materials, each labeled as a metal, insulator, or metal-insulator transition compound, based on available experimental literature.
At the time of this publication, there are 96 metals, 179 insulators and 68 MIT materials in the dataset.
\rev{Next, we obtained a crystal structure for each material via one of the following methods, in order of descending preference: retrieval from experimental library (ICSD or Springer), retrieval from the Materials Project, or in-house generation, as described in the SI.\cite{code-link}} 
The crystal structure of each material was then automatically featurized using common descriptors from \texttt{Magpie},\cite{Ward2016AMaterials}and those obtained from domain knowledge.

We labeled materials as MIT compounds if experimental literature on them shows that they exhibit an insulating ${\partial \rho}/{\partial T}<0$ temperature-dependent resistivity on one side of a critical transition temperature $T_\mathrm{MIT}$, and a metallic ${\partial \rho}/{\partial T}>0$ temperature-dependent resistivity on the other side of $T_\mathrm{MIT}$.
When there was ambiguity regarding the change in sign of 
the experimental ${\partial \rho}(T)/{\partial T}>0$ data, we used additional experimental data to determine the class label. 
For example, if optical data shows a finite charge gap at temperatures below  $T_\mathrm{MIT}$, but none above it 
(as is the case with the MIT in Sr$_3$Fe$_2$O$_7$ \citep{SFO}), we assigned an MIT label. 
Such subtleties are common among transition metal compounds, which are also prone to non-stoichiometry which can also influence class assignment, especially among less studied materials.

Readers can view the class-label assignments for each material at Ref.\ \citen{code-link}. %\url{https://github.com/MTD-group/mit_model_code}. 
This electronic database hosts the latest experimental 
information on thermal MIT compounds and related metals and insulators. 
The data we present herein represents the state-of-knowledge on the 
class labels for materials available at the time of publication.

\begin{figure}[t]
    \centering
    \includegraphics[width=0.45\textwidth]{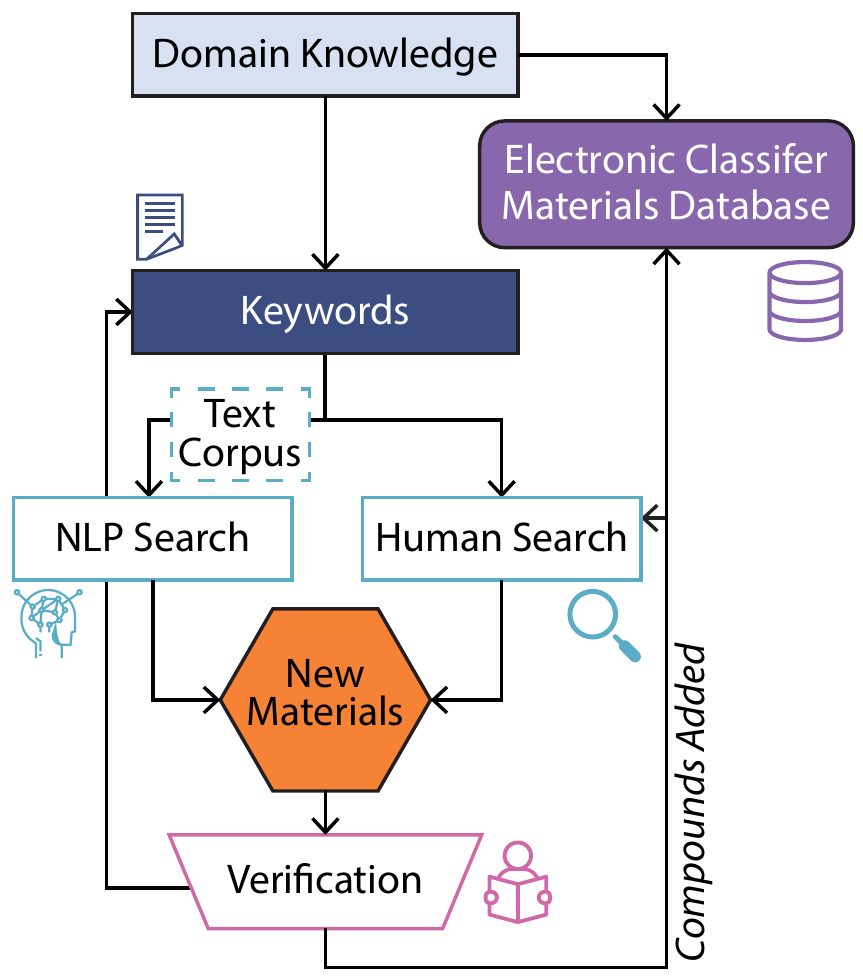}
    \caption{Iterative workflow, comprising natural language processing (NLP) and human searches, utilized to construct the database of materials for training the electronic classifiers.
    }
    \label{fig:database_workflow}
\end{figure}

\subsection{Data Acquisition and Database Construction}

\autoref{fig:database_workflow} illustrates the process by which we built the database for training the electronic classification models. 
We used a combination of domain knowledge and natural language processing (NLP) to generate both an initial materials database and a keyword list for the NLP pipeline. 
The materials database initially included all MIT compounds known by the authors along with related materials, including binary vanadates  V$_n$O$_m$, $R$NiO$_3$ nickelates ($R$ a rare-earth metal), LaCoO$_3$, and some  Ruddlesden-Popper oxides (e.g.,  Ca$_2$RuO$_4$), as well as lacunar spinels.
Compounds with similar chemistries and stoichiometries that are exclusively metallic or insulating were then also added to the database.

To extend the database beyond the materials known to the authors, we have also used natural language procesing (NLP). The text corpus used for the NLP methods included 70,123 papers, which was down selected from an entire text corpus of over 4 million scientific papers and journals. 
Down selection to the highly specialized MIT text corpus 
was performed using keywords (\supt1), describing MITs and correlated electron systems from the authors' domain knowledge, 
which matched words in the titles, abstracts, and introductory paragraphs of papers. 
New MIT compounds identified from two types of NLP searches (\emph{vide infra}) were then verified manually by assessing experimental transport (and/or optical) data.
The newly identified compounds were each assigned a metal, insulator, or MIT class label and added to the database. 
Based on these identifications, new keywords were also added 
to guide the NLP search or additional human searches to find more 
compounds relevant to the classification tasks.
This process was repeated until the database acquired 343 unique entries. 
%%

%\subsection{Natural Language Processing}
Two NLP methods were used in our workflow.
First, using the specialized MIT text corpus of approximately 70,000 papers and a state-of-the-art NLP pipeline,\cite{NLP1} we extracted chemical formulas of possible MIT compounds from each paper. 
This entity recognition process included tokenization of the relevant MIT text corpus, normalization to lemmatize each term, part-of-speech tagging using dependency parsers, and finally token classification. 
Using this NLP method, we identified the MIT compound PrRu$_4$P$_{12}$, 
which belongs to the skutterudite family frequently studied in 
thermoelectrics research.
Its transition is attributed to the physics of Pr 4f electrons, rather than the Ru 4d electrons.\cite{PhysRevLett.79.3218} 
Although the authors were unaware of this compound prior to the NLP search, 
we were able to combine our MIT domain knowledge to perform a manual search of the literature and identified SmRu$_4$P$_{12}$ as another skutterudite MIT material, as well as a wide range of other skutterudite materials which exhibit solely metallic or insulating behavior.

Second, we employed a FastText model trained directly on the specialized MIT text corpus.
The resulting word embeddings were then compared in terms of cosine similarity to previously identified MIT materials. 
We use a different cosine-similarity approach than that in Ref.\ \citen{NLP2} to identify compounds of interest.
For each compound in the original dataset of identified materials, 
100 words with highest cosine similarity in the trained FastText model were identified. 
Then using the metal, insulator, and MIT class labels present within the dataset, we grouped the closest 100 words for each compound into closest words for a given label. 
Because there were approximately 50 temperature-driven MITs in the original dataset, this grouping method resulted in $\approx$ 5,000 compounds with repeats, that were closest in cosine similarity of word embeddings to known MIT compounds.
Then within the group of compounds for a given class label, 
the 20 most commonly occurring words were identified, resulting in 
20 words for each classification label (60 words in total). 
Of the 60 words, the vast majority were chemical formulas with no 
noise.\footnote{Nine words were not exact chemical formulas of compounds and thus were considered to be noise. Although we assigned the words as noise, some words were still relevant; for example, 
the word ``La$A$O$_3$'' appeared in the 20 most common words associated with insulators, where $A$ represented an unidentified element on the periodic table.}
We further simplified the search by keeping only those words which were exact chemical formulas.  
After these compounds were identified, abstracts and titles were searched for these specific compounds. 
Among this subset of literature, we then searched the classifications for 
the compounds identified by the FastText model and 
added new materials to the database. 
The two NLP-assisted searches led to the addition of 116 compounds to the database (representing a 51\,\% increase in size from the initial dataset), which was then augmented further by additional human searches using the new domain knowledge. \rev{Our database was expanded from about 190 unique compounds to the current 343 with the addition of the mixed human + NLP search.
This expansion led to identification, for example, of the skutterudite family in which the MIT mechanism is driven by the rare-earth cations rather than transition metal-driven physics characteristic of the other materials in our database.}

\begin{figure}[t]
    \centering
    \includegraphics[width=\textwidth]{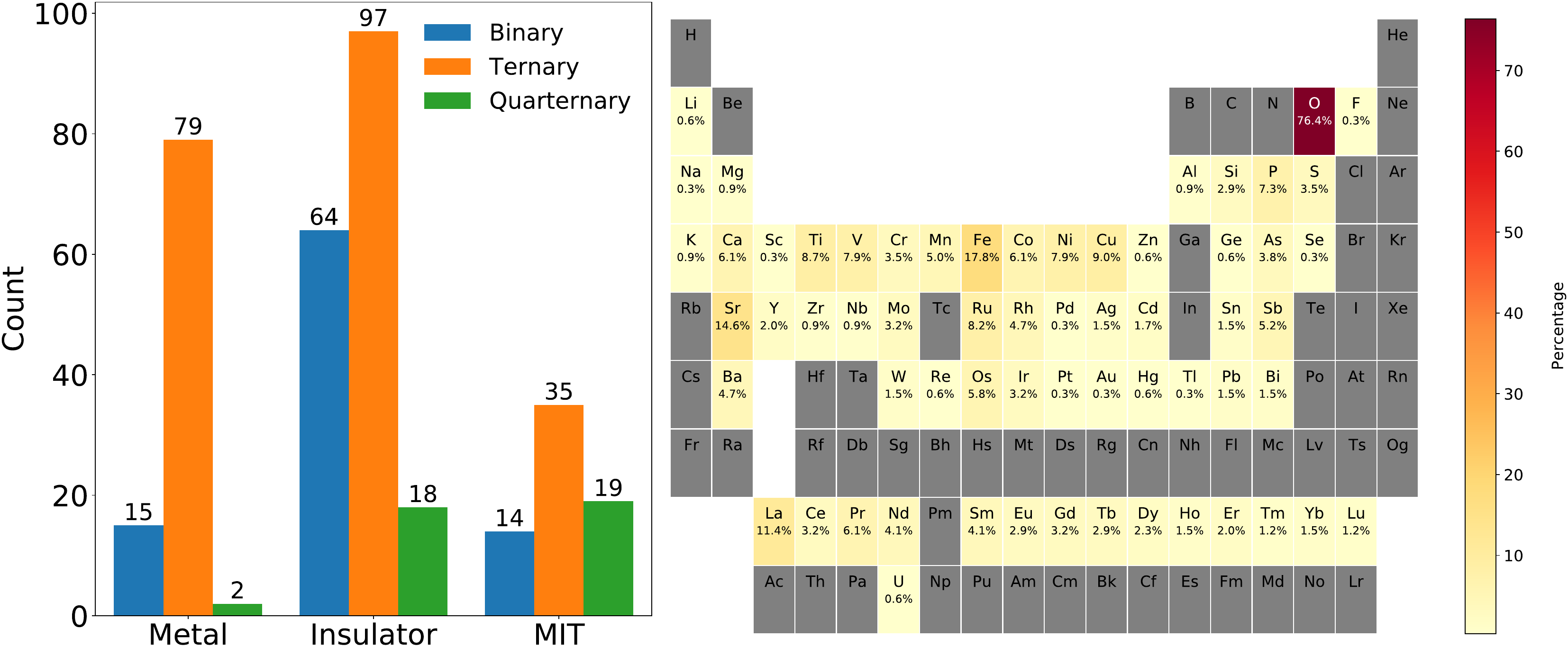}
    \caption{\rev{The distribution plot (left) shows the number of compounds by conductivity class. The heatmap plot of the periodic table (right) shows the elements that are present in all compounds in the database, and the value below each element symbol is the percentage of compounds that exhibit the corresponding element.}
    }
    \label{fig:class_distribution}
\end{figure}

As a result of this search, we have built a database that is as representative as possible. \autoref{fig:class_distribution} shows the distribution of the complete dataset obtained from this workflow at the date of submission of this paper. As expected for a database of materials dominated by transition metal oxides, the dataset is imbalanced, with the number of insulators significantly greater than that of metals or MIT materials.
All known MITs fall into a relatively restricted class of compounds and 
most frequently appear as complex ternary materials, often transition 
metal oxides and sulfides. 
In order for our database to be focused on thermal MITs, we have excluded compositionally-controlled MIT compounds. Our model then focuses on complex, inorganic materials involving d-shell or f-shell electrons as the valence states.
Although an arbitrary increase in our database size to go beyond this limited scope may allow for an increase in our classifier scores, this increase would not be meaningful as we want the metals and insulators to have similar chemical composition and stoichiometry to the MIT compounds. 
\rev{The heatmap in \autoref{fig:class_distribution} can be used as a quick guide to identify if a new compound to be tested contains elements that are already in the training data. This would better inform the decision as to how much trust one could place on the classification provided by our ML classifiers. Analogous heatmaps for metals, insulators or the MIT compounds can be found in the SI.}

As most MITs are accompanied by a structural change coupled to the electronic transition, an important question in building our model was which structure to select for featurization: the structure corresponding to the low-temperature, often insulating phase, or the high-temperature, often, metallic phase, or both? 
Across the electronic transition, a symmetry lowering distortion typically 
occurs with the low-temperature phase comprising small atomic displacements absent 
in the high temperature structure (see local distortions illustrated in 
\autoref{fig:mechanisms}). 
For example, the rare-earth nickelates at low-temperature have a breathing distortion, i.e., small and large  NiO$_6$ octahedra alternate in a 3D checkerboard pattern.\cite{PhysRevB.88.054101,Wagner2018}
This type of microscopic model associated symmetry-breaking is usually known only in materials after the compound has been studied sufficiently to be labeled as an MIT material. 
Furthermore, most crystal structures are initially reported at room temperature, which can be far from $T_\mathrm{MIT}$.
To build a model that is both mechanism-agnostic and can predict whether a material will have an MIT based on a simple theoretical or experimental structure before in-depth analysis could be performed, we opted to include only the high temperature structures in our dataset, if available. 
This allows our model to learn
the susceptibility of a compound to undergo an MIT using more readily available high-temperature structures. 

\subsection{Feature Engineering}
Constructing the machine learning model requires tabulating appropriate features to describe its properties for accurate class predictions. 
Our features include the \texttt{Magpie}\cite{Ward2016AMaterials} composition feature set, oxidation state, Ewald energy, local and crystal structure parameters (i.e., variation in bond lengths and atomic volumes), and the global instability index (GII)\cite{Salinas-Sanchez1992a}, all of which are accessible from the \texttt{Matminer}\cite{Ward2018} package.
Certain descriptors known from the material physics community to  be important in describing MIT compounds, however, are unavailable in these 
standard libraries designed  for machine learning.
To that end, we constructed  additional structural features, intended to capture the displacive distortions shown in \autoref{fig:mechanisms}, as well as built featurizers that can provide an estimate of the
electronic energy scales used in the Zaanen-Sawatzky-Allen framework\cite{PhysRevLett.55.418}
to separate metals from insulators.\cite{Torrance1991}

\begin{figure}[t]
\centering
\includegraphics[width=0.8\textwidth]{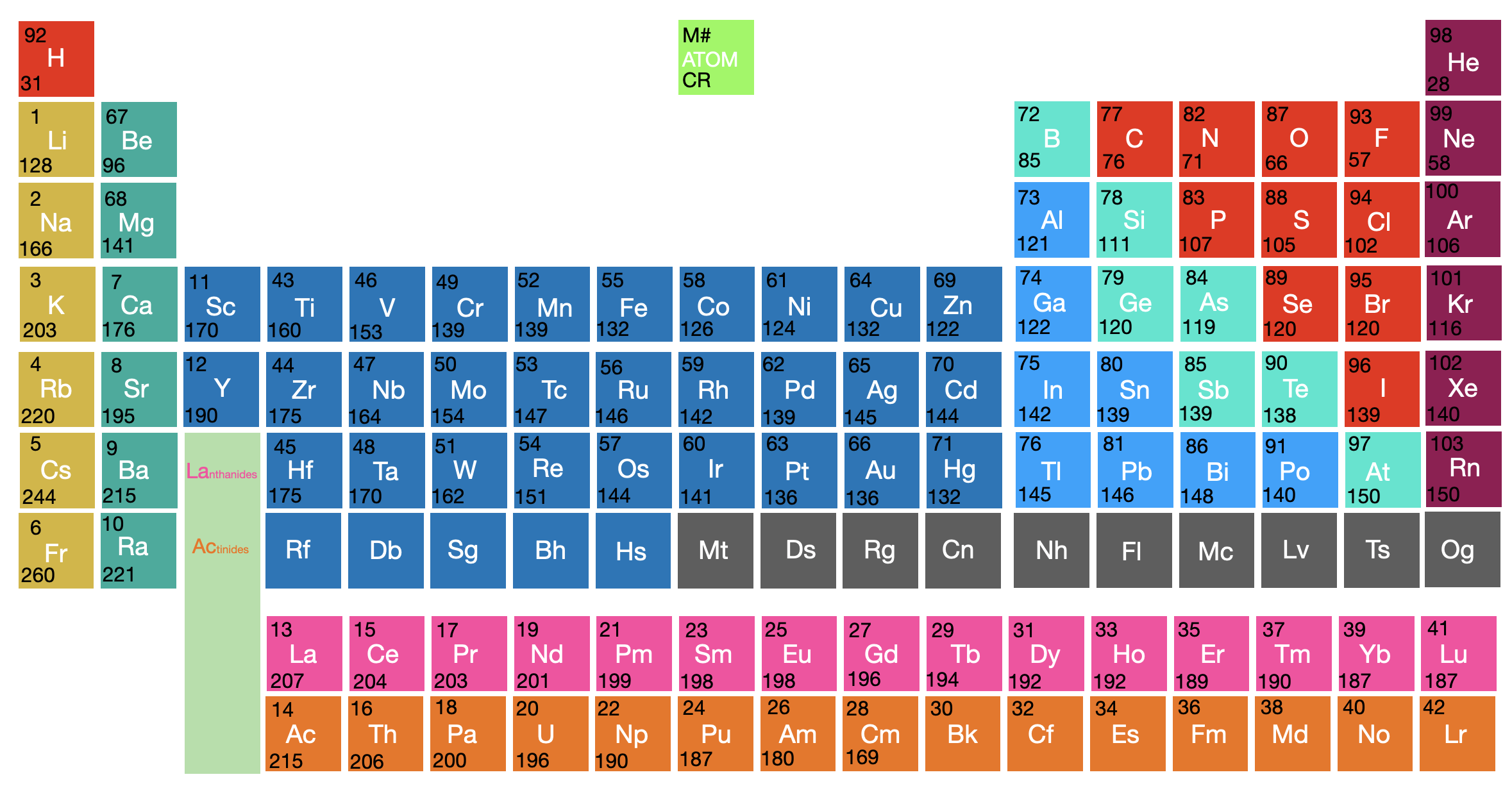}
\caption{Periodic table of Mendeleev numbers (M\#) and covalent radii (CR, in picometers) used for featurization.}
\label{fig:periodictable}
\end{figure}

Now, we will describe some of the features determined to be important and their implementation.
We begin by highlighting the range (or minimum) of the Mendeleev number and the average deviation of the covalent radius (see \autoref{fig:periodictable} for values used in this work). 
These two features both appear consistently as features with high importance in several training iterations, and have physically interpretable meanings. 
The Mendeleev number provides an alternative label beyond atomic number to distinguish elements with shared characteristics\cite{Villars2004}. The Mendeleev number generally (but now always) increases down the columns of the periodic table, then increases from left to right. This ordering is intended to bunch elements with similar chemical properties together by an expected oxidation state in most materials.
To understand how this can be useful, consider the $AB$O$_3$ perovskite oxides with $A$ a rare-earth and $B$ a transition metal.
The minimum of the Mendeleev number will characterize the $A$ cation and the maximum of the Mendeleev number will characterize the O anion. Thus, the range of the Mendeleev number will be the difference between the Mendeleev numbers of the O anion and the $A$ cation. 
Since most compounds in our dataset are oxides with just a few being sulfides, the maximum of the Mendeleev number is effectively fixed for a substantial portion of our dataset with only the minimum of the Mendeleev number varying between different compounds.
This aspect leads to high correlation between the minimum and range Mendeleev number features: 
the range and minimum of the Mendeleev number have a linear correlation of $-0.995$ and can be considered equivalent features for most of our dataset.
We chose to use the range of the Mendeleev number simply as it takes the minimum of the Mendeleev number into account, which may be important for other types of compounds (such as sulfides), and the effect of choosing one over the other is negligible on the performance of our models.

The average deviation of the covalent radius (ADCR) describes how different the covalent radii are for different elements in a compound:
\begin{equation}
    \label{eq:ADCR}
    \mathrm{ADCR} = \frac{1}{N}\sum_{i=1}^{N}|R_i-\bar{R}|,
\end{equation}
where $R_i$ is the covalent radius of the element $i$, $\bar{R}$ is the average covalent radius of all elements in a specific compound, and $N$ is the total number of different elements. 
For example in an $A_n$X$_m$ compound, 
\begin{equation}
    \mathrm{ADCR}\,({A}_n{X}_m)=\frac{n|R_A-\bar{R}|+m|R_X-\bar{R}|}{n+m}
\end{equation}
with the weighted average $\bar{R}=({nR_A+mR_X})/({n+m})$.
We use covalent rather than ionic radii as features, because while ionic radii are known to underlie structural stability in ionic crystals according to Pauling's rules,\cite{George2020} they rely upon knowledge of oxidation states and coordination environment. Further, some of our compounds exhibit non-integer oxidation states, or have oxidation states that are the subject of ongoing research (including many of the skutterudites), which means that including them would have introduced additional ambiguity into our model.

Now, we turn to some of the features for which we have built our own featurizers.
We begin with the structural
global-instability-index (GII)  descriptor, which is defined as
\begin{equation}
    \label{eq:GII}
    \mathrm{GII} = \left({\frac{1}{N}\sum_{i=1}^{N}d_i^2}\right)^{1/2},
\end{equation}
where $d_{i} = \mathrm{BVS}(i) - V_{i}$ is the difference between the bond valence sum (BVS) for the $i^{th}$ ion and its formal valence and $N$ is the number of ions in the unit cell.\cite{Brown1992,GII1992}
The GII is the root-mean-square deviation of the bond valence sums from the formal valence  averaged over all atoms in the cell.
This can be understood as approximating an average structural stress inherent in a material, as it captures the average deviation in bond lengths from what would be experimentally expected.
The stresses arise from  a combination of over- and underbonded cation-ligand interactions and thus describe bond strains compatible with a given structure and crystallographic symmetry.\cite{Brown/Poeppelmeier:2014} 

Next, we constructed additional structural features, which in turn allowed us to build the electronic features from the Zaanen-Sawatzky-Allen (ZSA) framework. The structural features include: 
the minimum, maximum, and mean distances between 
transition metal $M$ cations;
the minimum, maximum, and mean distances between the transition metal 
and the ligand $L$; and the 
Madelung site potentials for the cations and anions.
These features were then used to calculate approximations\cite{Torrance1991} for the 
relevant ZSA energy scales: the 
difference between the highest occupied and lowest unoccupied metal 
orbitals (an estimated Hubbard $U$, hereafter $U_0^\prime$)  
and a charge-transfer energy gap ($\Delta_0$).   
Both electronic parameters originate from an ionic model involving local charge excitations with 
\begin{equation}
    \label{eq:Uprime}
    U^\prime_0 = I_{v+1}(M)-I_{v}(M)-e^2/d_{M-M}\,,
\end{equation}
where $I_v=A$ is the electron affinity of $M^{v+}$, $I_{v+1}$ is the ionization potential, and $e^2/d_{M-M}$ is the Coulomb attraction between the excited electron on one transition metal cation and the hole left behind on its nearest-neighbor cation. The estimated charge-transfer energy is 
\begin{equation}
    \label{eq:Delta}
    \Delta_0 = e\Delta{}V_M+I(L^{n-})-I_{v}(M)-e^2/d_{M-L}\,,
\end{equation}
where $e\Delta{}V_M$ is the product of the electron charge and the difference in electrostatic 
Madelung site potentials between the cation and ligand sites, $I(L^{n-})$ is the ionization potential of the ligand in the $-n$ oxidation state, e.g.,   $I(\mathrm{O}^{2-}) = -7.7$\,eV for oxygen ligands. 
$I_v$ is as before and $e^2/d_{M-L}$ is the Coulomb attraction between the excited electron on the metal and the hole left behind on its ligand. 
Code for calculating $U^\prime_0$ and $\Delta_0$ was adapted from Ref.\ \citen{Hong2016} and is available in Ref.\ \citen{code-link}. The ionization potentials and electron affinity values were web-scraped from the NIST Atomic Spectra Database\cite{nist-atomic-database}.

\subsection{Supervised Learning Scheme}

\subsubsection{Machine Learning Algorithm}
To determine which machine learning model is best suited for our classification task, six different models were used in the model selection process: 
dummy classifiers with random guessing, 
linear logistic regression models with L2 regularization, 
generic decision tree models, 
random forest classifiers, 
gradient-boosting classifiers, 
and extreme gradient-boosting classifiers as implemented in XGBoost.\cite{Chen2016}
Model hyperparameters were optimized using grid search on the training split in stratified 5-fold cross-validation and are available at Ref.\ \citen{code-link}. 
%\url{https://github.com/MTD-group/mit_model_code#21-tune-the-xgboost-model}.

%
Tree-based ensemble methods have  been demonstrated empirically to be very efficient machine-learning models that also provide interpretability,\cite{Olson2018,Ward2018} %Jain2019} 
making it easier to expand our domain knowledge. 
Indeed, XGBoost is an optimized distributed gradient boosting library designed to be highly efficient, flexible and portable. 
We found the XGBoost models were consistently among the best performing models and 
they were relatively fast to train compared to random forest and 
gradient-boosting models as described in the SI. The accessibility of our code allows the user to test their own structures in a browser via the Binder service which we will discuss below or on their personal computer.
For these reasons, all classifiers presented here are based on XGBoost models, which are trained on two different features sets: a \emph{full feature set} and a \emph{reduced feature set} as described next.

\subsubsection{Feature Selection}
In order to obtain an easily interpretable model, and to avoid possible overfitting due to the large number of features, we performed a downselection to certain key features. This selection follows an iterative approach. In the first iteration, our raw feature set included 164 features (163 numeric features and 1 one-hot-encoded categorical feature with 2 levels), which is large compared to the number of compounds. To reduce the feature space complexity, we first removed any numeric features with 0 variance or with an absolute value of linear correlation greater than 0.95 with other features. This resulted in 106 features and is referred to as the \emph{full feature set}. 

Principal component analysis (PCA), t-distributed stochastic neighbor embedding (t-SNE), and uniform manifold approximation and projection (UMAP)\cite{McInnes2020} are often used to reduce the number of features. 
The linear and/or nonlinear combination of the original features used in these approaches to create new features, however, makes it difficult to interpret the physical meaning of each new descriptor. 
Since we desired to preserve the physical meaning of our features, we used a combination of Shapley additive explanations (SHAP)\cite{lundberg_unified_2017} and domain knowledge (physical intuition) to downselect the features in the second iteration. 
For each of the three binary classifiers, SHAP analysis on the full feature set was used to find the 
10 most important features, i.e., the top 10 features with the highest average absolute SHAP values. 
From this SHAP analysis, 6 features were selected and then combined with 4 features chosen using domain knowledge, which resulted in a total of 10 features. 
This second feature set is referred to as the \emph{reduced feature set}. 
Both the full and reduced feature sets are available at Ref.\ \citen{code-link}.
%\url{https://github.com/MTD-group/mit_model_code/tree/master/data/processed/csv_version}. 

We also note that the SHAP scores of each feature are highly dependent on the training dataset: minor updates to the dataset result in significant changes in SHAP importance scores, leading to the conclusion that these measures may not be reliable if taken individually. This is likely a result of our small dataset: in the full feature set, we have 343 compounds and 106 features to describe them, leading to potential overfitting, which will be further addressed in future versions of the code. 
As a result, we handpicked a combination of features that either appear consistently as important in our SHAP analysis or we believed to be relevant from physical intuition.
\rev{For instance, as we increased the number of compounds and trained models on the different iterations of the dataset, we found several features that consistently exhibited high SHAP values, such as the global instability index and the average deviation of the covalent radius. These features were then combined with those deemed important from materials domain knowledge such as the average metal-metal distance, the Hubbard $U$ strength, and charge transfer energy, to form the reduced feature set. The physical interpretation of these features is discussed in more detail below.}

\subsubsection{Model Metrics}

We computed model classification performance metrics such as receiver operating characteristic (ROC) curves and precision-recall curves with stratified cross-validation splits. 
Because the splits are dependent on the random seed used to generate them, and performance can vary depending on the different train-validation splits from different seeds, we performed cross-validation using 10 random seeds with integers from 0 to 9. 
For each of the 10 seeds, we performed a stratified 5-fold cross validation from which we calculated a median value.
All metric values we report hereafter are the median values along with the interquartile range of those 10 median values.  
We carried out all of our cross-validations with the \texttt{scikit-learn} \cite{scikit-learn} Python package. 
Weighted F-1 scores that take class-imbalance into account were also used for model assessment.

\rev{For a dataset this small, a test set usually should not be} used to evaluate model performance as there are only 343 training examples; it was unfeasible to set aside a hold-out set, since to the best of our knowledge, our current temperature-driven MIT materials database is exhaustive. However we will address this issue in future versions of our code, particularly as our database expands.
Through cross-validation on 10 random seeds, each with stratified 5-fold splits, we then use the cross-validation performance as a proxy for the actual test set approach. \rev{Nonetheless, we did include for completeness a 90\%-10\% train test split evaluation using the same 10 random seeds, which resulted in similar performance, as reported in the SI. The key difference here is that for the original cross-validation approach, the hyperparameter tuning process uses the entire dataset while during the train test split, it uses only 90\% of the data, and thus the performance is more indicative of the models' extrapolative power.}

\section{Results}

\subsection{Classifier Performance}
We first present performance results for our three binary classifier models, as they perform significantly better than a single ternary classifier (see \supf1 of SI).
Classifier M distinguishes between metals and non-metals, which include compounds labeled as insulators (I) or MIT compounds (T). 
Classifier I is the analogous classifier for insulators with non-insulators comprised of M and T compounds.
Classifier T distinguishes MIT compounds from non-MIT compounds, i.e., metals and 
insulators. 

\begin{figure}
    \centering
    \includegraphics[width=0.45\textwidth]{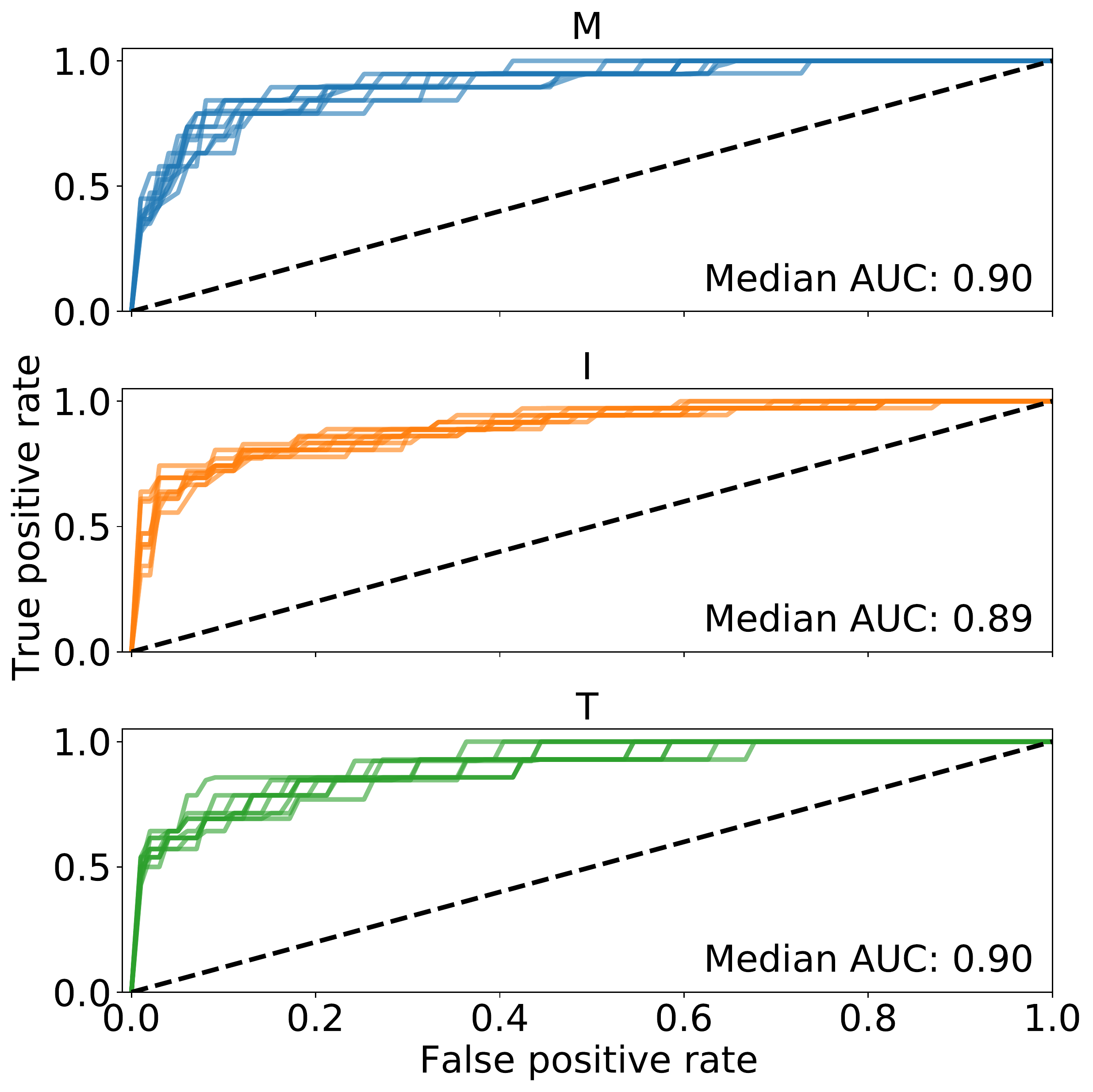}
    \caption{Receiver operating characteristic curves for the binary metal (M), insulator (I), and MIT compound (T) classifiers. Each colored line represents the median ROC curve from one of the 10 random seeds. For each seed, a 5-fold stratified cross-validation is carried out. Dashed lines represent the performance ($\text{AUC}=0.5$) when randomly guessing. The median area under the curve (AUC) is provided in the lower right corner.}
    \label{fig:AUC}
\end{figure}

\begin{figure}
    \centering
    \includegraphics[width=0.45\textwidth]{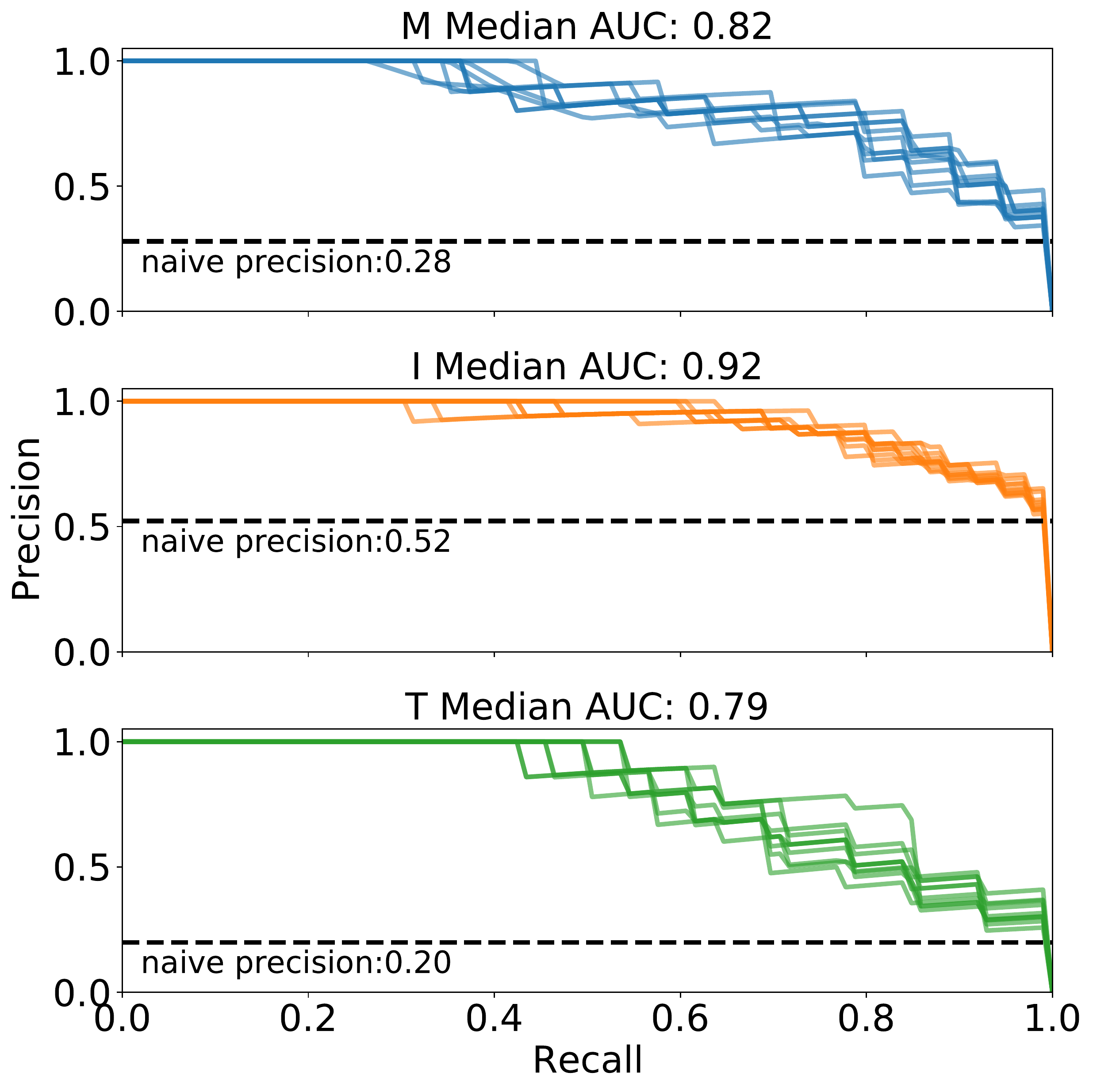}
    \caption{Precision recall curves from 10 seeds with each seed having a 5-fold cross-validation for the binary metal (M), insulator (I), and MIT compound (T) classifiers. The naive precision represented by the black dashed lines indicate what the precision would be if a classifier always predict the positive class. Here ``AUC" is the area under the precision-recall curve, not to be confused with the area under the receiver operating characteristic curve in \autoref{fig:AUC}.
    }
    \label{fig:prec-recall}
\end{figure}

All model performance metrics presented are from models trained on the 
reduced feature set. The corresponding metrics from models trained on the full feature set are available at Ref.\ \citen{code-link}. 
Except for the I classifier, which has a slightly worse performance for the one trained on the reduced feature set than that trained on the full feature set, the M and T classifiers are able to retain the same level of performance when trained on the reduced feature set compared to those trained on the full feature set (see \supf2).

To visualize model performances, ROC curves for our machine learning models are constructed across 5 random cross-validation runs (\autoref{fig:AUC}) under 10 random seeds.
The ROC curve depicts the true positive rate, which is the ratio between the number of correctly identified positive class (true-positives) and all of the positive class (the sum of true-positives and false-negatives), against the false-positive rate, or the ratio of false-positives to the sum of false-positives and true-negatives. 
An ROC curve confined primarily to the upper left corner is an indication of a model correctly identifying instances of each class without many false positives.  
An area under the ROC curve (AUC) of 1 represents perfect separation whereas an AUC of 0.5 is equivalent to random guessing.
Tighter bunching of the lines indicates less variance in model performance 
with varying random seeds. We obtain a median ROC area under curve (ROC-AUC) of 0.90 and 0.89 with interquartile ranges of 0.02 and 0.01 for the M and I classifiers, respectively. 
Remarkably, our novel T classifier exhibits a median ROC-AUC of 0.90 with an  interquartile range of 0.03, indicating its overall accuracy is high.

\autoref{fig:prec-recall} presents the precision (proportion of true positives to the sum of true positives and false positives) and recall (proportion of true positives to the sum of true positives and false negatives) curves of each binary classifier to better understand performance owing to imbalance among the classes.
The median and interquartile range of the cross-validation weighted F$_1$ scores (harmonic mean of precision and recall that takes class imbalance into account) are 0.86\,(0.03), 0.82\,(0.02), and 0.88\,(0.01) for the M, I, and T classifiers, respectively. 

On one hand, we find that the area under the precision-recall curve for Classifier T is rather low. 
Indeed, MITs are the least represented class in the dataset owing to the small number of known thermally-driven MIT materials.
The poor precision-recall performance could perhaps be overcome with additional 
data as seen in other works (\supt2).
As more positive examples (MIT compounds) are added to the dataset, we expect the precision of Classifier T to improve because the under-representation of MITs in the training set may lead to under-prediction of MITs, which results in a smaller number of true positives and thus a lower precision. 
On the other hand, the performance of Classifier M and I is comparatively better than that of Classifier T since the dataset contains more metals and insulators. As a result, the models were able to better separate metals from non-metals and insulators from non-insulators. In other words, these two classifiers exhibit better performance, because for them the ratio of positive class to negative class is more balanced than that of Classifier T. 

Compared with electronic state classifiers formulated in earlier works using various databases and different descriptors (\supt2), our models' performance metrics for the M and I models are comparable. However, we note that the previous models were not intended to learn correlated electron materials, and their metrics if trained on our sparse dataset would likely be different. Therefore, the comparison is not strictly appropriate.

We also created a survey that let domain experts (e.g., materials scientists) classify the conductivity class of 18 compounds (6 metals, 7 insulators and 5 MIT compounds). The goal was to establish a human performance baseline for the 3 aforementioned classifiers and to evaluate whether identifying MIT compounds is a trivial task for human scientists. Unsurprisingly, the XGBoost classifiers outperform the average human scientist in every classification task (see \supf3 of SI).

\subsection{Feature Importance and Physical Interpretation}

\begin{figure}[t]
    \centering
        \includegraphics[width=0.75\linewidth]{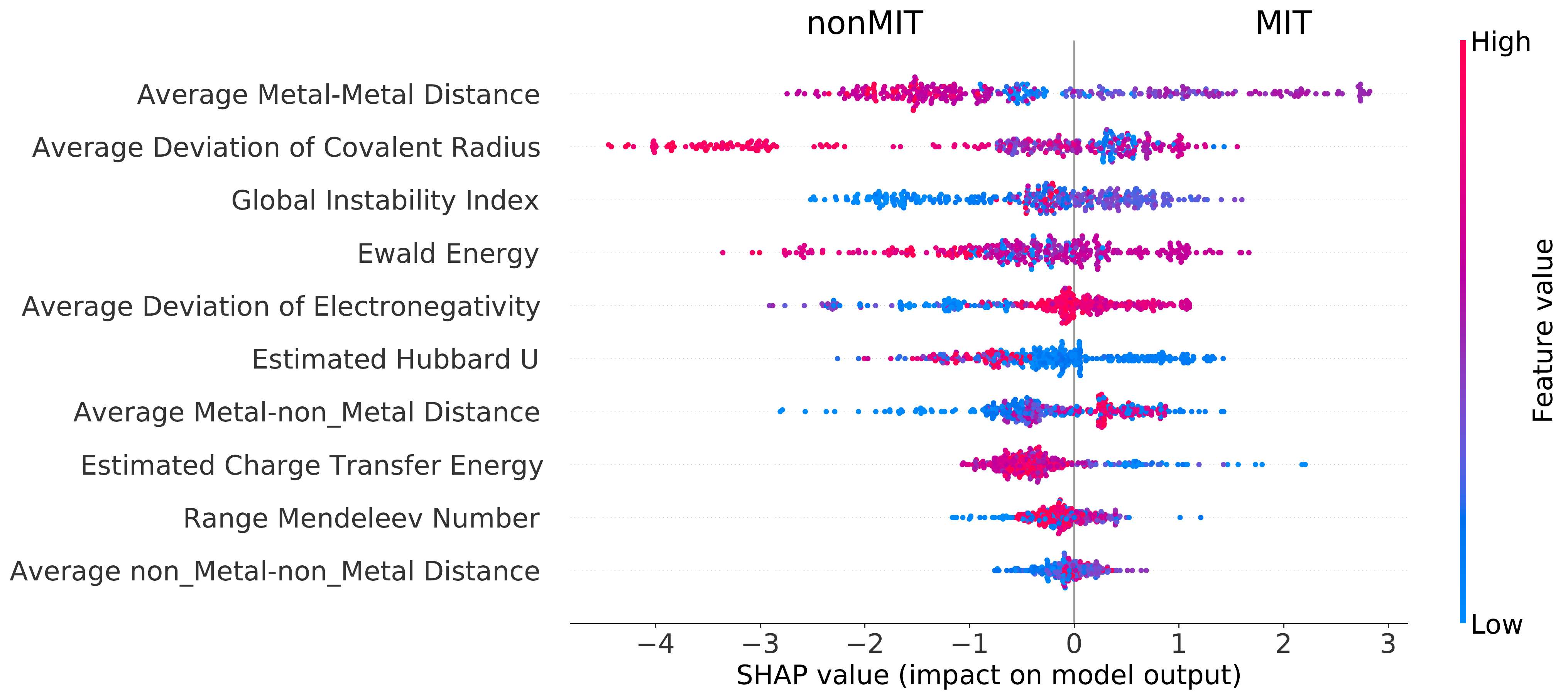}
    \caption{SHAP feature importances for the MIT compound (T) Classifier from the reduced feature set. For each row of feature, each dot corresponds to one compound in the database (i.e., every row has 343 dots). The vertical location of a feature name reflects its importance as ranked by mean absolute SHAP value (the higher up the more important). Colors of each dot indicate whether the corresponding feature was of high (red) or low (blue) value for that compound in the dataset relative to the maximum and minimum values of the feature for all materials in the database. The horizontal location of each dot is the SHAP value, which indicates whether that feature value causes a higher or lower prediction probability of the material as an MIT compound by log-odds. Dots are displaced vertically to reflect the density of compounds at a given SHAP value.}
    \label{fig:MIT_importances}
\end{figure}

We now use a combination of domain knowledge, SHAP values, and Accumulated Local Effect (ALE) analysis 
to examine the role of different features in the T Classifier trained on the reduced feature set. 
We want to know how the model learns to differentiate an MIT compound from those that are exclusively metallic or insulating. 
SHAP values indicate which features are important and the effects of the features on the classification (i.e., how does changing the value of a feature change the classification). 
ALE plots play a similar role in elucidating  
the importance of each feature in classifying a material, as well as the role of the \emph{interactions} between the features. Some of the most illustrative ALE plots can be found in the SI.
\autoref{fig:MIT_importances} shows the rank-order SHAP importance of each feature as well each feature's relative role in the MIT compound classification (e.g., whether a material having a small value of GII makes it more likely to be classified as an MIT compound or not).

\begin{figure}[h]
    \centering
    \includegraphics[width=0.70\textwidth]{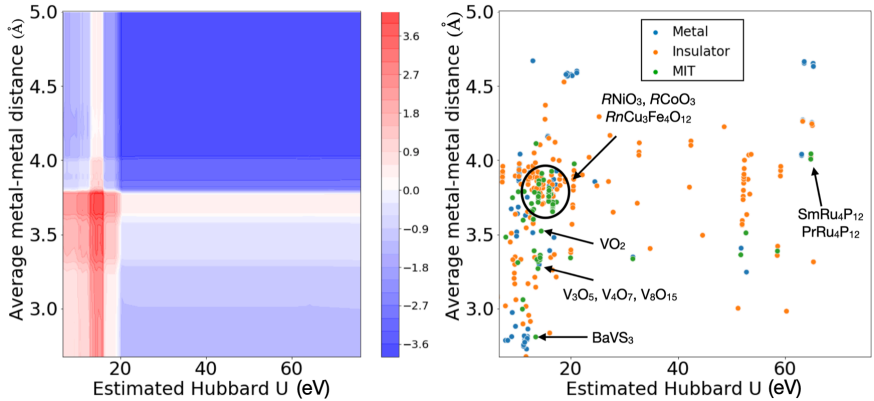}\vspace{-48pt}
    \caption{Interplay of the distance between identical transition metal ions and the unscreened Hubbard $U$ interaction. The ALE plot (left) shows the contribution to the classification probability from these two features, with a higher value (red) corresponding to a higher probability of a positive MIT classification. The scatter plot (right) shows the distribution of compounds in our dataset as a function of these two features, with select families labeled.
    We note that most of the MITs that are part of the ABO$_3$ perovskite family, and many of the vanadate MIT compounds are close to the ideal region for these two features. MIT materials such as skutterudite SmRu$_4$P$_{12}$ and PrRu$_4$P$_{12}$ are not considered by the model to have MITs driven by Hubbard $U$ or the metal-metal distance.}
    \label{fig:mmU}
\end{figure}

\begin{figure}[h]
    \centering
    \includegraphics[width=0.45\textwidth]{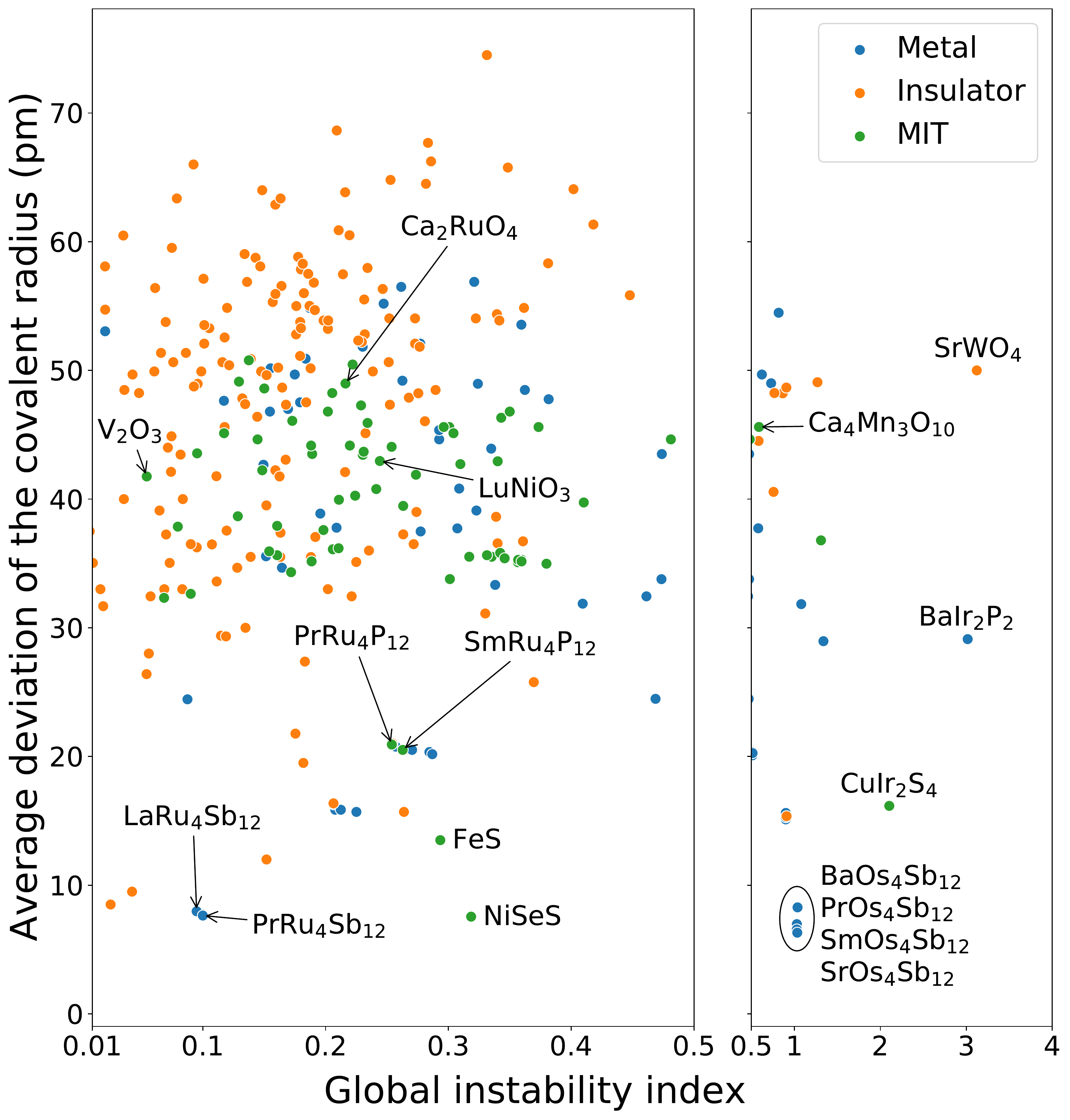}
    \caption{Interplay of GII (abscissa) and ADCR (ordinate). Most MIT compounds exhibit intermediate ADCR and GII values with low ADCR exceptions for FeS, NiSeS, CuIr$_2$S$_4$. %
        An interactive version of the data in this feature plot is available at {\small\url{https://mtd.mccormick.northwestern.edu/mit-classification-dataset}.}
    }
    \label{fig:CovRadius-GII}
\end{figure}

The most important feature in our classification is the average transition metal - transition metal distance, which we refer to as metal-metal distance for brevity, as shown in Fig. \ref{fig:MIT_importances}. The ALE decomposition shows that the importance of this feature largely arises  from its interaction with other features. This finding agrees well with our physical intuition. In transition metal compounds, the metal-metal distance may determine the electronic bandwidth $W$, which is in competition with other energy scales  such as the Hubbard $U$, as described for example in the ZSA-classification scheme, that drive MITs. 
Indeed, we show in a 2D scatter plot the distribution of materials as a function of the metal-metal distance and Hubbard $U$ and the ALE plot for these two features in Fig. \ref{fig:mmU}. 
We find that most of the most well-studied MIT materials (mostly perovskite compounds) fall within a narrow region identified as high in probability based on these two features. Interestingly, for some materials for which the mechanism is largely unknown, (SmRu$_4$P$_{12}$ and PrRu$_4$P$_{12}$), but  widely assumed to be different than that of most other MIT materials, these two features do not contribute to the MIT classification in any significant amount. This agrees with previous theoretical work, which has suggested that the relevant MIT physics for these materials is, in fact, not driven by the transition metal ion electrons at all, and is driven by the Pr and Sm f-electrons instead.\citep{skutheory}

The SHAP feature importances further find the relevant features for the classification are the GII, the charge transfer energy, and the transition metal-ligand distances (see \supf10 with SHAP plots for these two compounds in the SI), providing a possible hint as to the relevant physics in these materials.    
The GII and ADCR are two features that we find to be consistently important, and novel, in our model. The GII has previously been related to MIT temperatures in certain materials families, for example in the the $Rn$Cu$_3$Fe$_4$O$_{12}$. \citep{giicufe}
\autoref{fig:CovRadius-GII} shows most MIT compounds exhibit ADCR values between $30\,\mathrm{pm}$ and $50\,\mathrm{pm}$ and GII values between 0.1 and 0.5. 
The moderate to high GII values for most MIT materials are consistent with our understanding that 
these thermally-driven MITs are assisted by a minor structural instability, which can alleviate bond stresses. Materials with high  GII values may be too chemically unstable to support this type of mechanism.
We find that the MIT compound with the lowest GII is V$_2$O$_3$. 
Overall, a low GII tends to favor an insulating state, and a higher GII favors a metallic or MIT state. 
For example, among binary oxides with rutile structure and composition $M$O$_2$, we 
find 
$\mathrm{GII\,(TiO}_2)=0.11$, 
$\mathrm{GII\,(VO}_2)=0.13$, and 
$\mathrm{GII\,(MoO}_2)=0.32$. 
As most of our compounds are oxides, and most of those are insulators with a low GII, 
we deduce that materials with a low GII are highly stable from a bond-stress assessment 
and are unlikely MIT compounds, which is consistent with the GII SHAP data in 
\autoref{fig:MIT_importances}.

\begin{figure}[t]
    \centering
    \includegraphics[width=0.45\textwidth]{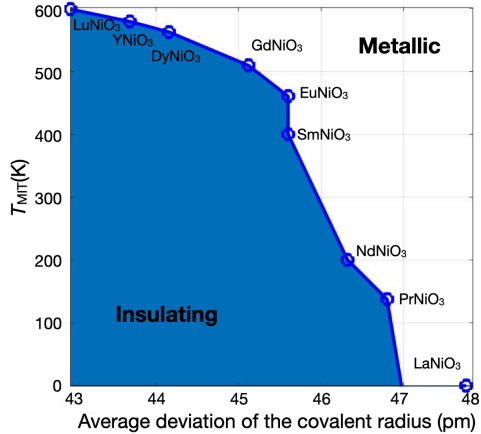}
    \caption{Phase diagram with $T_{\mathrm{MIT}}$ for RNiO$_3$ compounds versus the average deviation of the covalent Radius for the perovskite nickelates. 
    LaNiO$_3$ is always metallic. }
    \label{fig:nickelate}
\end{figure}

\begin{figure*}[t]
    \centering
    \includegraphics[width=0.98\textwidth]{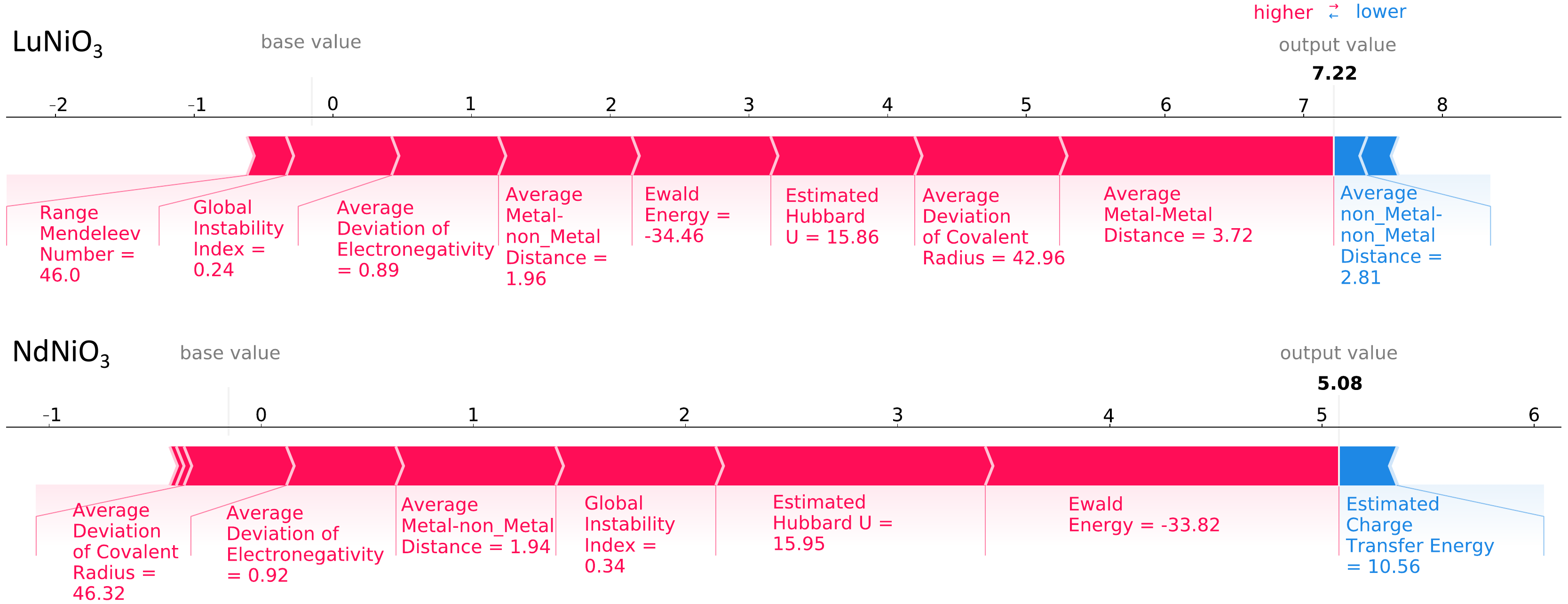}%\vspace{-60pt}
    \caption{SHAP force plot of predictions from Classifier T for the MIT compounds LuNiO$_3$ and NdNiO$_{3}$. Color indicates whether the feature taking the particular value was providing evidence for (red) or against (blue) a prediction of having an MIT. Bar size represents the SHAP value. SHAP values from all features sum up to the log-odds of a positive MIT prediction. The base value is the log-odds expected based on the average proportion of MITs in the dataset. The classifier predicts with higher confidence that LuNiO$_3$ is an MIT material than NdNiO$_3$, with the ADCR's contribution to the classification as extracted from Shapley values of 1.05 and 0.48, respectively.}
    \label{fig:SHAP-example}
\end{figure*}

\begin{figure}[t]
    \centering
    \includegraphics[width=0.45\textwidth]{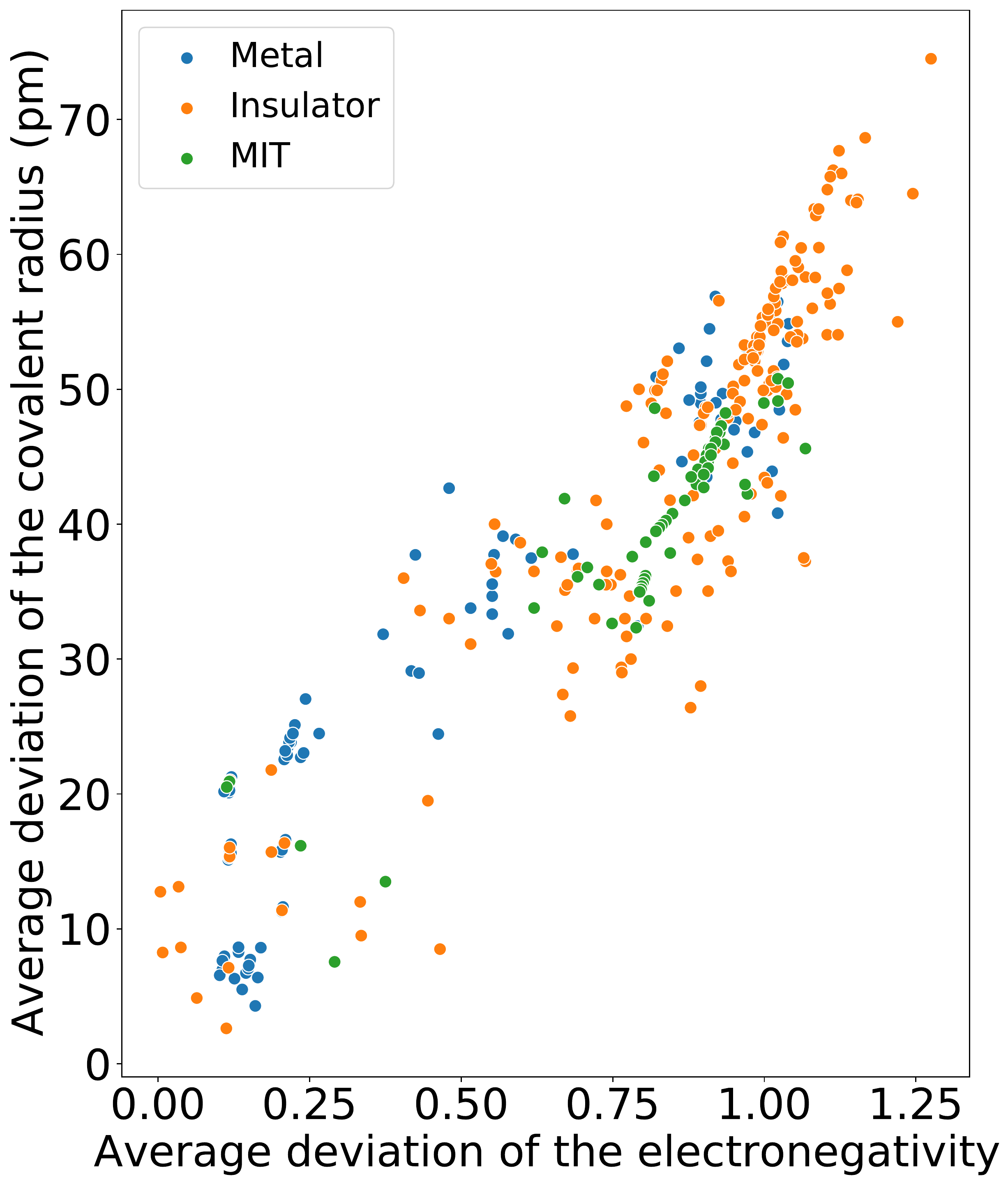}
    \caption{2D scatterplot of the average deviation of the covalent radius and the average deviation of the electronegativity features. Although the two features are strongly linearly correlated across our materials database, the MIT materials nonetheless show strong bunching.
    }
    \label{fig:elec-covrad}
\end{figure}

We glean some  understanding of the ADCR relevance by focusing on 
the perovskite $R$NiO$_3$ nickelates with $R$ a rare-earth cation.\footnote{A similar analysis can also be done on other perovskite families such as $R$CoO$_3$, or $Rn$Cu$_3$Fe$_4$O$_{12}$, with $R$ defined as before and $Rn$ further includes Ho, Tb, Tm.}
In cubic perovskites, ABO$_3$, including the RNiO$_3$, the ADCR is linearly correlated to the Goldschmidt tolerance factor ${t}=({r_R+r_\mathrm{O}})/(\sqrt{2}(r_B+r_\mathrm{O}))$. This tolerance factor is known to be associated with whether $B$O$_6$ octahedral rotations are likely to occur and distort the ideal cubic perovskite structure ($t=1$)\cite{Bartel2019}.
For $t<1$, the transition metal-oxygen octahedra rotate, making it more difficult for electrons to hop and favoring an MIT. Thus, a lower tolerance factor usually leads to a higher MIT temperature, while a higher $t$ can suppress MIT behavior altogether (e.g., LaNiO$_3$ is metallic). 
The ADCR plays a more important role in supporting the MIT classification for LuNiO$_3$ (lower ADCR) than it does for NdNiO$_3$ (higher ADCR), capturing the physical trend in the phase diagram in \autoref{fig:nickelate}: the ADCR places NdNiO$_3$ close to metallic LaNiO$_3$, while it places LuNiO$_3$ significantly further away towards high $T_{\mathrm{MIT}}$.
This physics is captured in the SHAP values in \autoref{fig:SHAP-example}: 
NdNiO$_3$  is one of the nickelates with the highest ADCR and the lowest $T_{\mathrm{MIT}}$, making it 
close to the metallic class as identified by the model with a log-odds ratio of 5.08.
In contrast, LuNiO$_3$ with an ADCR lower than  NdNiO$_3$ has a 7.22 log-odds ratio. 
In other words, the classifier is more certain that LuNiO$_3$ is an MIT compound than it is for NdNiO$_3$.
The ADCR is thus similar to a generalized tolerance factor, irrespective of the materials family studied. We note that the ADCR is also strongly correlated to the average deviation of electronegativity with a linear correlation of 0.919 (see \autoref{fig:elec-covrad}),
which we understand as a consequence of the electron affinity of an element being  partially determined by its atomic radius.

\begin{figure}[t]
\centering
    \includegraphics[width=0.45\textwidth]{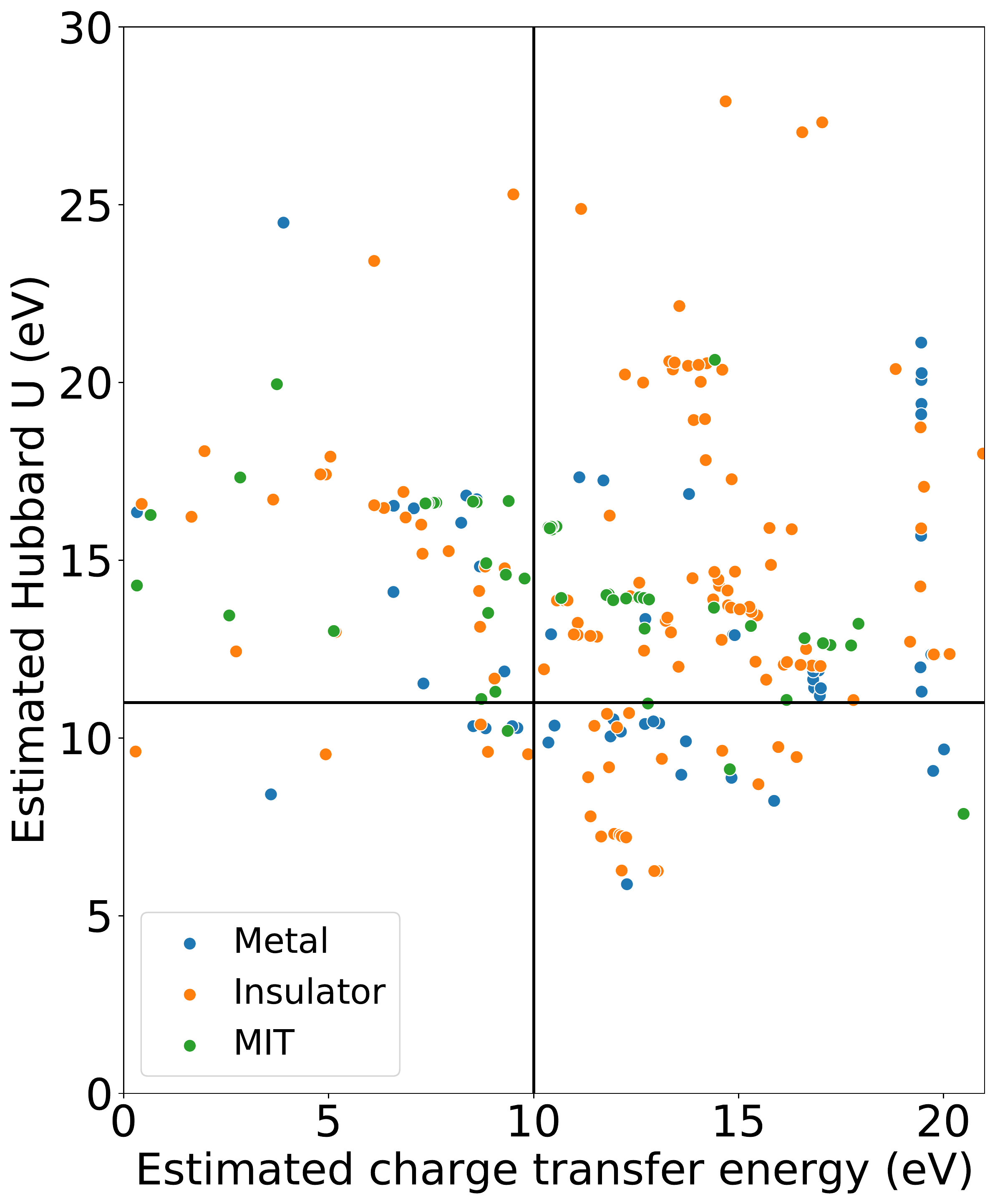}
    \caption{Our database of materials plotted in the two-dimensional space of 
    $\Delta_0$ (abscissa) and $U_0^\prime$ (ordinate) after 
    Torrance {et al.} \cite{Torrance1991} reveals poor class separation.
    The black lines at $\Delta_0 = 10$\,eV and $U^\prime_0 = 11$\,eV correspond to the boundaries between insulators (located in the upper right quadrant) and metals elsewhere as determined in Ref.\ \citen{Torrance1991}. MITs should lie close to the separation boundaries.}
    \label{fig:torrance}
\end{figure}

Although we find the estimated Hubbard $U$ values $U_0^\prime$ and 
charge-transfer energies $\Delta_0$, which are important to the ZSA classification 
of correlated metals and insulators, are among the top 8 features, 
the MIT materials do not strongly cluster when they are plotted in a 2D space consisting of these 
features (\autoref{fig:torrance}).
The GII, ADCR, and the range of the Mendeleev number (discussed later) lead to much stronger class clustering (or separation) than the ZSA-classification energies $U_0^\prime$ and $\Delta_0$, which have been used extensively over the last 30 years.
The Hubbard $U$ is a strong counter-indicator of an MIT when large and somewhat less strongly predictive of an MIT when low, which gives some support to the findings in Ref. \citen{Torrance1991}.
The presence of materials with extremely high  $U_0^\prime$  values arising from 
high ionization energies, e.g., titanates, distorts the SHAP color scale for the 
Estimated Hubbard $U$ row in \autoref{fig:MIT_importances}.
High values of $\Delta_0$ may also indicate that no MIT occurs, although sometimes the SHAP value of such a material is close to 0. The color scale for the charge transfer energy is also skewed by the presence of negative $\Delta_0$ values. These occur when the difference in metal and anion Madelung site potentials is small and/or the ionization energy of the metal is large.
One reason that the ZSA classification energies may be difficult to interpret is that the 
energy 
estimates we use correspond to unscreened values. 
Dielectric screening and metal-ligand hybridization 
effects in solids can lead to significant renormalization of these values, but require electronic-structure based 
calculations such as cRPA \citep{cRPA} to ascertain.  
Thus, although these numbers  provide some information to our machine learning model, they are difficult to understand in isolation.

Two additional features of high importance include the average metal-metal distance 
and the average metal-nonmetal (metal-ligand) distance.
We understand their role in relation to the energies $U_0^\prime$ and 
$\Delta_0$ within the ZSA framework, which are often scaled relative to the electron-hopping 
parameters describing correlated electron materials in the form of the d-orbital bandwidth in transition metal compounds. 
As the d-orbital bandwidth in the low-energy electronic structure 
is not directly available from the structure alone, 
a convenient proxy is the metal-metal distance and/or the metal-anion distance as 
the bandwidth is inversely proportional to the distance between the atomic pairs contributing states that hybridize by symmetry. This explains the strong role of inter-atomic distances in our model.

\begin{figure}[t]
    \centering
    \includegraphics[width=0.45\textwidth]{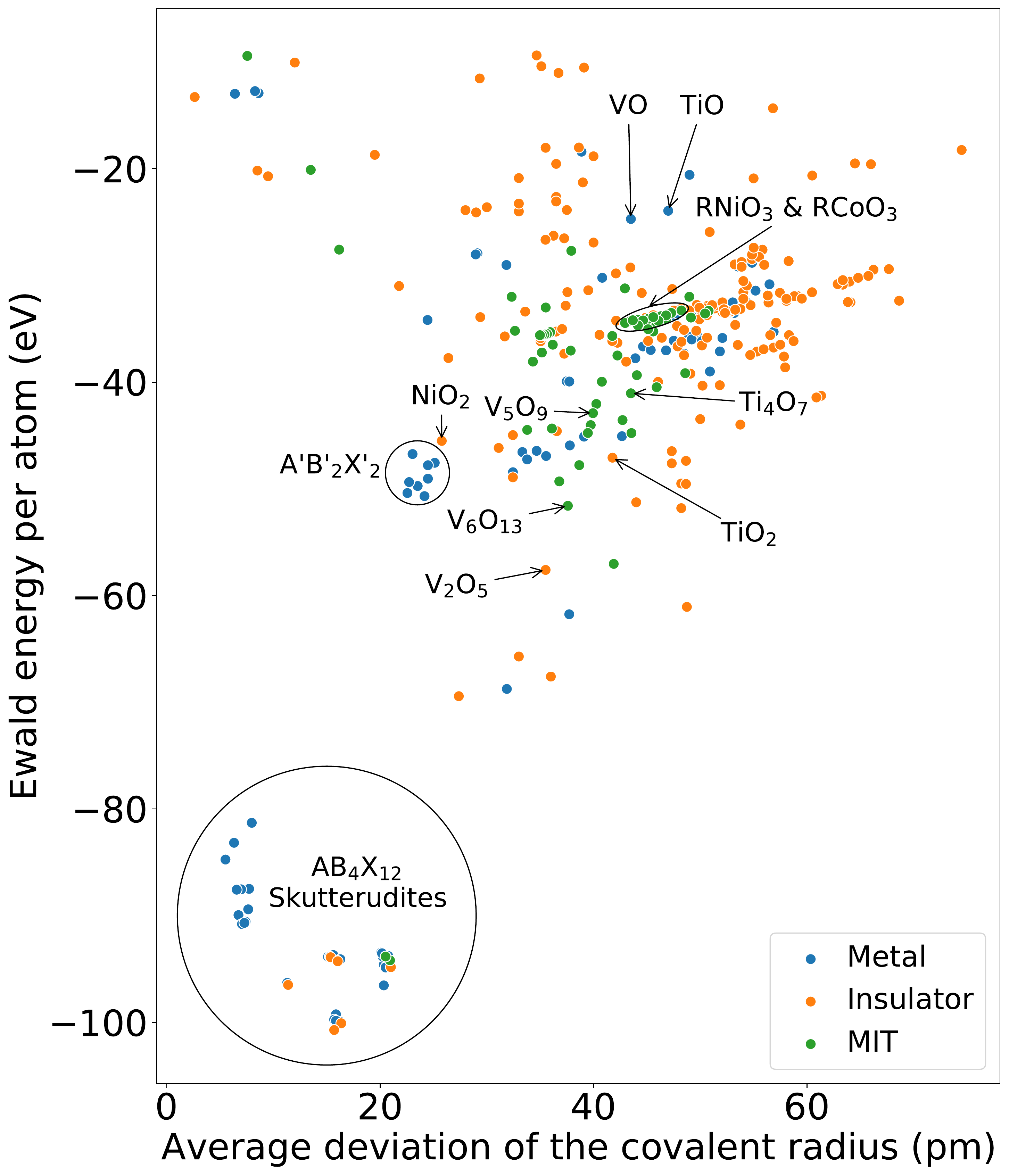}
    \caption{Interplay of Ewald energy per atom and ADCR. Note the linear correlation within the same family (i.e., binary Ti and V oxides) and the clustering of the compounds based on their anion. 
    $R$ and $Rn$ and yttrium are rare-earth metals, with the same valence, as defined in the main text.
    }
    \label{fig:CovRadius-Ewald}
\end{figure}

The Ewald energy reflects how stabilizing the ionic charge distribution in a crystal is due to the electrostatic potential imposed by oppositely charged ions in the atomic structure. 
The calculations as currently performed in the latest version of \texttt{Matminer} correspond to an electrostatic energy between ions modeled as point charges, with the charge approximated by the nominal ionic oxidation state. This feature, as implemented in \texttt{Matminer} has recently been updated to be normalized per atom.
This Ewald energy per atom then, to first order, separates highly ionic materials from less ionic materials. Phosphates such as CoP$_3$ (which has an Ewald energy of $-124\,\mathrm{eV/atom}$) have strong ionic character in this picture, with P having a $3-$ charge, and Co a $9+$ charge, while sulfides tend to have lower absolute values of the Ewald energy  (FeS$_2$ has an Ewald energy of $-10\,\mathrm{eV/atom}$). 
\autoref{fig:CovRadius-Ewald} shows oxides that exhibit intermediate values.

\begin{figure}[t]
    \centering
        \includegraphics[width=0.45\textwidth]{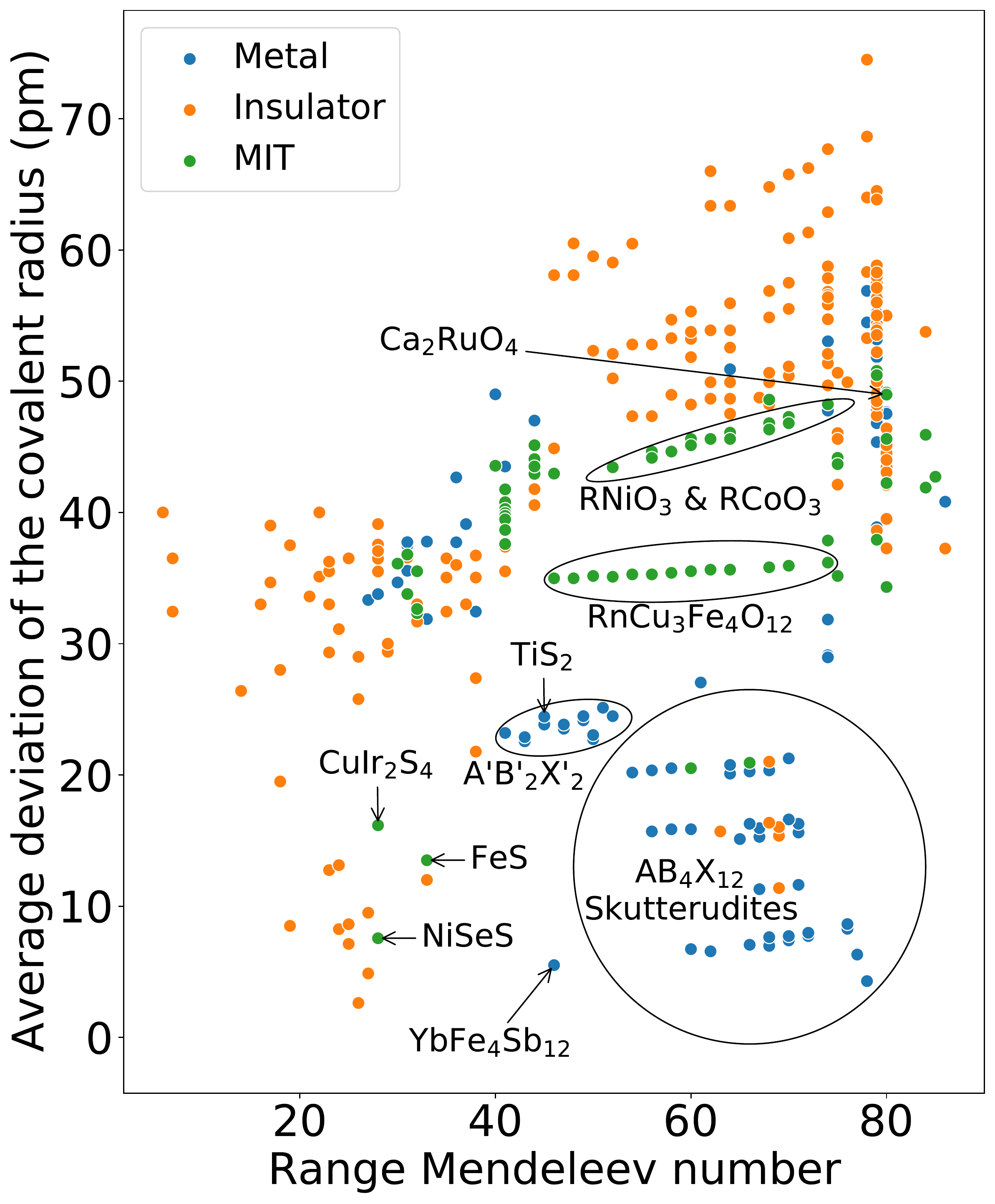}
    \caption{Interplay of ADCR and Range Mendeleev Number. Many MIT compounds exhibit an ADCR between 30 and 50 pm, and are comprised of group IIIB elements or lanthanides.  
    $R$ and $Rn$ and yttrium are rare-earth metals, with the same valence, as defined in the main text. RnTm$_4$X$_{12}$ skutterudites include an Rn element as defined before however including Ba, and Ce as well, Tm a transition metal ion from Fe, Ru or Os, and X one of the Sb, As or P anions. The RnTm$_2$X$_2$ are defined the same way as the RnTm$_4$X$_{12}$ skutterudite.}
    \label{fig:CovRadius-RangeMend}
\end{figure}

This separation based on the anion Ewald energy is similar to that based on the maximum Mendeleev number, which may explain why the maximum of the Mendeleev Number (which describes the anion in our compounds) does not have a role in our classification according to our SHAP scores whereas the range of the Mendeleev number is much more important. 
The range of the Mendeleev number essentially separates 
compounds that contain elements from the first three columns of the periodic table or from the lanthanide or actinide series (such as LaNiO$_3$ or EuO), from those that are binary transition metal compounds (such as FeS or NiO). The importance of this feature is clearer to discern through its interaction with other features, whereas the Ewald energy alone leads to a clear separation based on composition, particularly based on the anion type. %
From the combination of ADCR and the range of the Mendeleev number, we find strong clustering of MIT materials (\autoref{fig:CovRadius-RangeMend}). 
Particularly, a range of the Mendeleev number over 40 and an ADCR below 50 are likely an indicator of an MIT material. 
We also find that most known MIT compounds tend to be oxides containing yttrium or a lanthanide in their composition (as highlighted by the green dots enclosed in horizontal ellipses in \autoref{fig:CovRadius-RangeMend}).

Although the Ewald-energy-SHAP scores in \autoref{fig:MIT_importances} are difficult to interpret for the entire database as a whole, they are easier to understand if we focus on a particular family. 
Consider the V$_{n}$O$_{m}$ binary vanadium oxide family.
We find that V$_2$O$_5$ with an Ewald energy of
$-57\,\mathrm{eV/atom}$ is insulating, 
while VO with an Ewald energy of % -24.713
$-24\,\mathrm{eV/atom}$ is metallic.
The V$_6$O$_{13}$, VO$_2$, V$_8$O$_{15}$, V$_6$O$_{11}$, V$_5$O$_9$, V$_4$O$_7$, V$_3$O$_5$ and V$_2$O$_3$ 
vanadates exhibiting intermediate Ewald energies are all MIT compounds. 
This observation strongly suggests that the Ewald energy of the MIT compounds within a particular materials family is likely to lie in between that of the metallic and insulating members in that family.
A similar analysis can be performed for the Ti$_n$O$_m$ family (\autoref{fig:CovRadius-Ewald}).

The Ewald energy is less useful, however, for differentiating materials with the same 
stoichiometry and structure type but comprised of different chemistry: 
LaNiO$_3$ has an Ewald energy of 
% -33.532eV/at, 
$-33\,\mathrm{eV/atom}$ and LuNiO$_3$ has an Ewald energy of 
% -34.4603eV/at 
$-34\,\mathrm{eV/atom}$
despite the very large differences between the two as illustrated in 
\autoref{fig:nickelate}.
As the materials classes have varying ranges of Ewald energy, it is then difficult to understand its role from the SHAP plot alone without focusing on a particular family.

Finally, we also included the average deviation of the electronegativity as a feature for our classifier. Although it is strongly linearly correlated with the ADCR, we surprisingly find that there is a strong clustering of the MIT compounds within this 2D feature space (\autoref{fig:elec-covrad}).

\begin{figure}[t]
    \centering
    \includegraphics[width=0.65\textwidth]{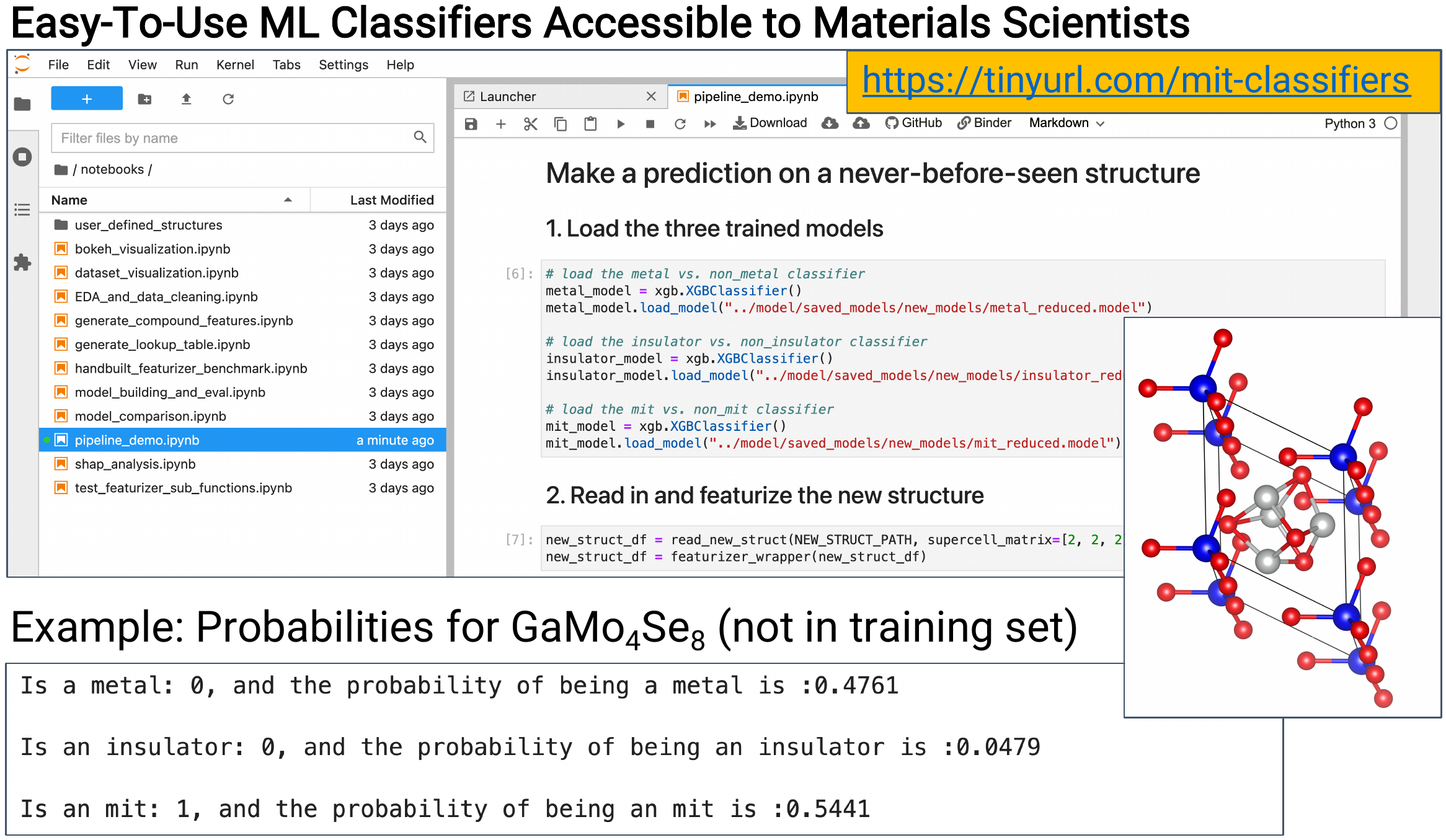}
    \caption{Screenshot of the web-based Binder notebook, which permits a user to submit a crystal information file (CIF) for a compound and obtain three electronic classification probabilities: 
    metal \emph{vs} non-metal (M), insulator \emph{vs} non-insulator (I), and MIT \emph{vs} non-MIT (T) as shown for GaMo$_4$Se$_8$.
    }
    \label{fig:binder-demo}
\end{figure}

\subsection{Online Classifiers for Predicting Conductivity Classes}
Our pre-trained electronic classifiers are deployed and served to the larger materials science community through a cloud service called Binder\cite{binder}. There are several Jupyter notebooks\cite{Kluyver2016jupyter} hosted on the Binder server that may be used to easily reproduce the results presented herein. 
The Binder website offers interactive execution of these notebooks directly in a web browser without installing any dependency onto a local machine. All notebooks are also available at Ref.\ \citen{code-link}. 
% \url{https://github.com/MTD-group/mit_model_code#demo-notebooks}. 

Here we present a brief demonstration of the Binder notebook at \url{https://tinyurl.com/mit-classifiers}, 
which enables a user to upload a structure in CIF format and immediately make a classification using our models, without the need to install any software. 
After uploading a lacunar spinel structure GaMo$_4$Se$_8$, which was identified computationally as a potential MIT material\cite{wang2020featureless},  the notebook automatically featurizes the new structure and 
makes a prediction. 
Then after executing through the three binary classifiers trained on the reduced feature set, 
GaMo$_4$Se$_8$ is classified as an MIT material (\autoref{fig:binder-demo}). 
Note among the three possible conductivity classes, the assigned classifications may not be mutually exclusive since a single ternary classification is not made. 
The classifiers predict the following probabilities: 0.4761 for being a metal; 0.0479 for being an insulator; 0.5441 for being an MIT compound. The default threshold for making a positive classification is 0.5, which means a positive classification is made only if the predicted probability is greater than 0.5. In this case, GaMo$_4$Se$_8$ is predicted only to be an MIT compound. However it is also worth noting that although GaMo$_4$Se$_8$ does not obtain  a positive metal classification, its probability for being a metal is 0.4761, which is close to 0.5.

\section{Conclusions}

Within this work, we highlighted and attempted to resolve two important issues in the field of metal-insulator transition compounds, which are in fact quite general to the study of quantum materials more broadly. First, the lack of a widely accessible database of materials based on a particularly relevant - but rare - property. And second, a methodology to provide insight into this class of materials that is complementary to that of standard electronic structure and model calculation methods. 

Our electronic materials database comprising of MIT compounds as well as related metals and insulators, will help broaden the domain knowledge of other scientists in the field. On this database, we trained three easy-to-interpret machine learning models. \rev{We recognized that the training data size is limited, and took measures to avoid over-fitting whenever possible. We offered a brief analysis on the robustness and extrapolation power of the MIT classifier in the SI.} Based on the MIT classifier model, we identified new features that determine whether a material has a temperature-driven MIT or not, advancing our domain knowledge for this type of classification problem. 
Particularly, we found the Global Instability Index, Average Deviation of the Covalent Radius, Ewald Energy, and 
Range of the Mendeleev Number as well as combinations between pairs of these features to be important to the performance of the MIT classification model.

\rev{The high importance of the transition metal-transition metal distance, as well as its interaction with other features, such as the Hubbard $U$, highlights both the ability of machine learning approaches to gain physical insight and confirms previous theories about the nature of the MIT in an unbiased way.}
%
%\sout {Domain-science descriptors based on estimated Hubbard interactions and charge transfer energies were found to be less important, however they were still needed for a higher quality model.} \rev{I don't think this is true anymore, especially in our new model and with ALE analysis}
%
MIT materials exhibited strong clustering when plotted in a 2D space spanned by two novel features, the average deviation of the covalent radius (ADCR) and the range of the Mendeleev number, making it possible to quickly assess whether novel materials discovered in the laboratory or predicted computationally may exhibit MITs. 
We also provided a periodic table with the Mendeleev number and covalent radius of the atoms 
to enable a quick calculation for other scientists of these two features.
We conjecture that these features may be relevant in creating simple models analogous to the  Goldhammer-Herzfeld criterion,\citep{GF} which allows for the differentiation between elemental metals 
and nonmetals.
Finally, we offered a simple-to-use online platform that allows users to upload a crystal structure file and obtain a probabilistic prediction on the electronic class of their material.

\begin{suppinfo}
The Supporting Information is available free of charge on the ACS Publications website at DOI: the performance comparison of XGBoost classifiers (1) against other types of machine learning models (e.g., random forest) trained on the full feature set, (2) trained on the full feature set against those trained on the reduced feature set, and (3) trained in this work against other models trained in previous works to classify only metals and insulators; short primer on SHAP values; NLP search keywords; survey analysis comparing classification accuracy of domain experts against XGBoost classifiers; ALE analysis; \rev{element heatmaps for metals, insulators and MIT compounds; model evaluation with holdout test sets; and a brief discussion on the robustness and extrapolation power of the MIT classifier.}
The materials database and calculated features are available online at Ref.\ \citen{code-link}.
\end{suppinfo}

\acknowledgement
The authors thank Professors R.\ Seshadri and S.\ Wilson at the University of California, Santa Barbara, for helpful discussions about this project.
This work was supported in part by the National Science Foundation (NSF) under award number DMR-1729303.
The information, data, or work presented herein was also funded in part by the Advanced Research Projects Agency-Energy (ARPA-E), U.S.\ Department of Energy, under Award Number DE-AR0001209.
The views and opinions of authors expressed herein do not necessarily state or reflect those of the United States Government or any agency thereof.
A.B.G.\ and P.R.\ contributed equally to this work, which was initiated by N.W.
A.B.G.\ identified the handpicked features, performed the physical analysis of the results, the human identification and classification of the materials in the final database, and helped coordinate the project.
P.R.\ built the final version of the classifier models, the online pipeline, and the featurizer used throughout the project.
A.R.T.\ built the NLP pipeline used and identified relevant compounds from the pipeline to add to the materials database.
S.Z.\ performed the ALE analysis.
K.M.\ built the webpage for the materials database.
D.A.\ supervised S.T.
E.A.O.\ supervised A.R.T.
J.M.R.\ conceived and administered the project.
All authors contributed to writing and revising the paper.

%\bibliography{references}

\providecommand{\latin}[1]{#1}
\makeatletter
\providecommand{\doi}
  {\begingroup\let\do\@makeother\dospecials
  \catcode`\{=1 \catcode`\}=2 \doi@aux}
\providecommand{\doi@aux}[1]{\endgroup\texttt{#1}}
\makeatother
\providecommand*\mcitethebibliography{\thebibliography}
\csname @ifundefined\endcsname{endmcitethebibliography}
  {\let\endmcitethebibliography\endthebibliography}{}

\newpage
\begin{figure*}
\center
\includegraphics[height=1.74in]{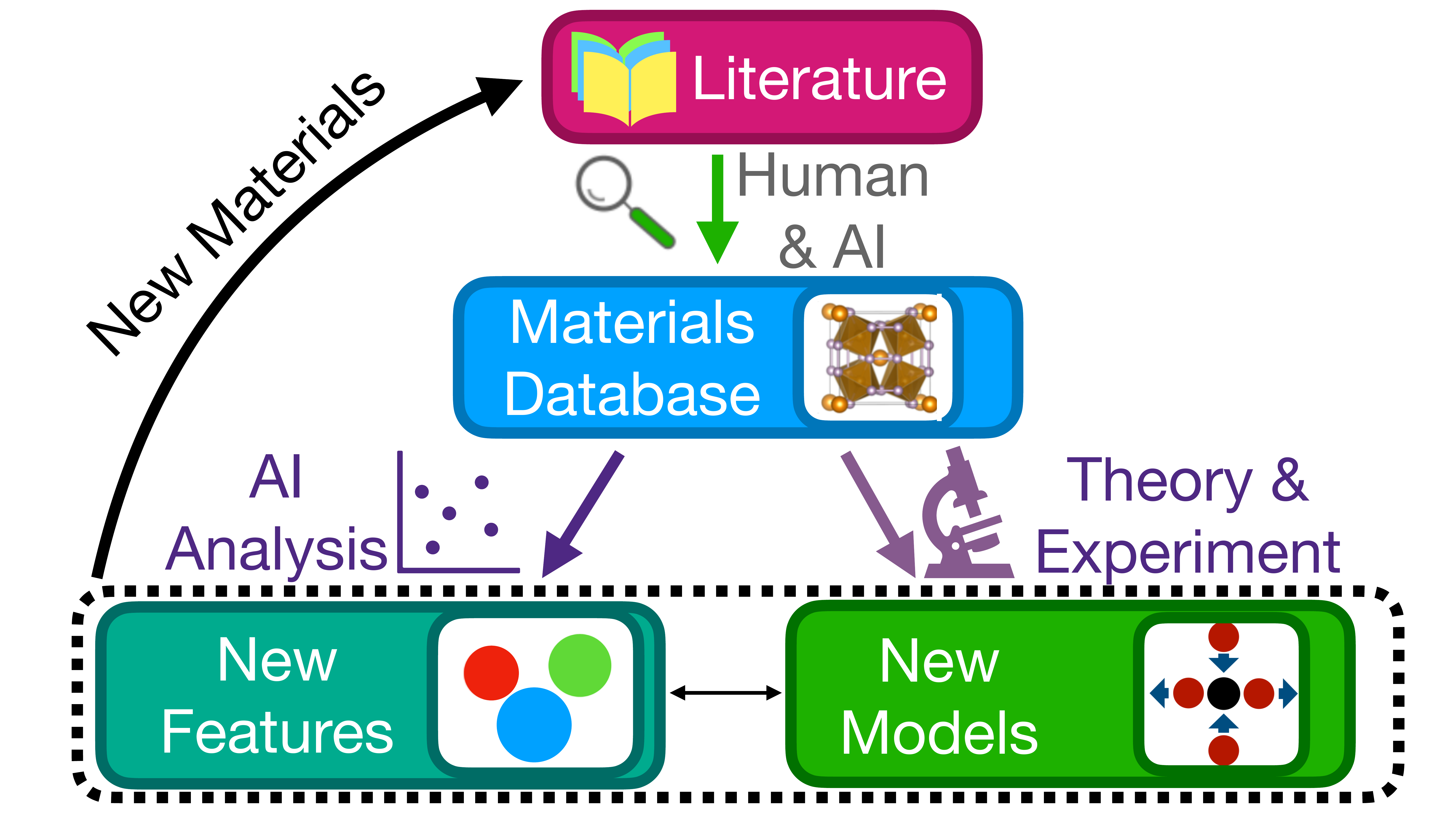}
%3.25 inches by 1.75 inches (approx. 8.25 cm by 4.45 cm). 
\end{figure*}
 \centering Table of Contents Graphic 
%\caption{Materials discovery and study workflow, integrating conventional theoretical and experimental study with database building and machine learning}

\end{document}